\documentclass[%
reprint,
twocolumn,
superscriptaddress,
showpacs,preprintnumbers,
nofootinbib,
aps,
prd,
]{revtex4-1}

\usepackage{epsfig}
\usepackage{amsfonts}
\PassOptionsToPackage{hyphens}{url}\usepackage[breaklinks,urlbordercolor={1 1 1}]{hyperref}

\usepackage[utf8]{inputenc}
\usepackage{ulem}
\usepackage{graphicx}
\usepackage{xcolor}
\graphicspath{ {./figures/} }

\usepackage[caption=false]{subfig}

\usepackage{comment}
\usepackage{amsmath}
\usepackage{bbold}
\usepackage{enumerate} 

\usepackage{multirow}

\newcommand\RE{\operatorname{Re}}

\newcommand\MSbar{$\overline{\text{MS}}$ } 
\newcommand\Veff{V_\text{eff}} 
\newcommand\he[1]{#1^\dagger}
\newcommand\gr[1]{\mathrm{#1}}
\newcommand\SU[1]{\mathrm{\gr{SU(#1)}}} 
\newcommand\grU[1]{\mathrm{\gr{U(#1)}}} 
\DeclareMathOperator{\Tr}{Tr}

\newcommand\sumint[1]{\int\kern-1.5em\sum\nolimits_{#1}}

\newcommand{\nn}{\nonumber \\}
\newcommand{\Tc}{T_{\rm c}}

\newcommand{\der}[2][]{\frac{d#1}{d#2}}

\begin{document}

\newcommand{\HEL}{\affiliation{%
        Department of Physics and Helsinki Institute of Physics,
        PL 64, FI-00014 University of Helsinki,
        Finland }}
	
\newcommand{\TSU}{\affiliation{Tsung-Dao Lee Institute \& School of Physics and Astronomy, Shanghai Jiao Tong University, Shanghai 200240, China }}

\newcommand{\AMH}{\affiliation{Amherst Center for Fundamental Interactions, Department of Physics,\\
University of Massachusetts, Amherst,
MA 01003, USA }}

\newcommand{\CAL}{\affiliation{Kellogg Radiation Laboratory, California Institute of Technology,\\
Pasadena,
CA 91125, USA}}

\newcommand{\TSUb}{\affiliation{Shanghai Key Laboratory for Particle Physics and Cosmology, Key Laboratory for Particle Astrophysics and Cosmology (MOE), Shanghai Jiao Tong University, Shanghai 200240, China }}
	
	\title{
        Nonperturbative study of the electroweak phase transition in the real scalar singlet extended Standard Model
    }
	
	\preprint{HIP-2024-7/TH}
    \preprint{ACFI T24-03}
	
	\author{Lauri Niemi} \email{lauri.b.niemi@helsinki.fi} \HEL \TSU
	\author{Michael J.~Ramsey-Musolf} \email{mjrm@sjtu.edu.cn} \TSU\TSUb\AMH\CAL
	\author{Guotao Xia} \email{xiagt-summer@sjtu.edu.cn} \TSU\TSUb

	\begin{abstract}
 We perform a nonperturbative lattice study of the electroweak phase transition in the real singlet scalar extension of the Standard Model. We consider both the heavy and light singlet-like scalar regimes at non-zero singlet-doublet mixing angle. After reviewing features of the lattice method relevant for phase transition studies, we analyze the dependence of phase transition thermodynamics on phenomenologically relevant parameters. In the heavy singlet-like scalar regime, we find that the transition is crossover for small doublet-singlet mixing angles, despite the presence of an energy barrier in the tree-level potential. The transition becomes first order for sufficiently large mixing angles. We find two-loop perturbation theory to agree closely with the lattice results for all thermodynamical quantities considered here (critical temperature, order parameter discontinuity, latent heat) when the transition is strongly first order. For the light singlet-like scalar regime relevant to exotic Higgs decays, we update previous one-loop perturbative results using the two-loop loop dimensionally reduced effective field theory and assess the nature of the transition with lattice simulations at set of benchmark parameter points. For fixed singlet-like scalar mass the transition becomes crossover when the magnitude of the Higgs-singlet portal coupling is small. We perform our simulations in the high-temperature effective theory, which we briefly review, and  present analytic expressions for the relevant lattice-continuum relations.
  
	\end{abstract}
	
	\maketitle
	
	\newpage

	\section{Introduction}

    A fundamental feature of the electroweak (EW) sector of the Standard Model (SM), and virtually all weakly-coupled extensions to it, is the Higgs mechanism: mass generation through condensation of a scalar field. This effect can be lifted, at high temperature, by thermal fluctuations.
    Determining the thermal history of this electroweak phase transition (EWPT) is a forefront challenge for particle physics and cosmology. While the Standard Model EW transition is known to be a smooth crossover and not a true phase transition for a 125 GeV Higgs boson \cite{Kajantie:1996mn, Csikor:1998eu}, an extended scalar sector can readily admit a first order electroweak phase transition (FOEWPT).

    Such a transition could have provided the necessary preconditions for generation of the cosmic baryon asymmetry via electroweak baryogenesis if the transition were sufficiently strong (see \cite{Morrissey:2012db} for a review). A strong FOEWPT could also have provided a source for a stochastic gravitational wave background that may be accessible with next generation probes such as LISA \cite{LISACosmologyWorkingGroup:2022jok}, Taiji, and Tianqin. Collider searches for beyond Standard Model (BSM) scalars, as well as precision measurements of Higgs boson properties, provide powerful and, generically, definitive probes of a strong FOEWPT \cite{Ramsey-Musolf:2019lsf}. 
	
	Extended scalar sectors continue to receive considerable attention, both theoretically and experimentally, and they span a wide range of scenarios, ranging from simplified model extensions to ultraviolet (UV) complete theories, supersymmetric and otherwise. For a recent review of these scenarios and their generic implications for a first order EWPT, see Ref.~\cite{Ramsey-Musolf:2019lsf}. The simplest extension consistent with a 125 GeV Higgs boson and a strong FOEWPT entails the addition of a real singlet scalar, the so-called \lq\lq xSM\rq\rq\ . 

    The xSM provides a useful simplified model stand in for scalar sector extensions that do not carry SM gauge charges ({\it e.g.} dark sectors). At the renormalizable level, the xSM contains two Higgs portal interactions whose couplings govern the thermal dynamics of EWPT. These interactions also carry distinct phenomenological consequences, making the xSM a rich arena for BSM collider physics. 
	
	In this work, we report on the first-ever non-perturbative study of thermal EWPT in the xSM in which the singlet scalar $S$ is retained as a dynamical degree of freedom. As a consequence of the well-known Linde problem \cite{Linde:1980ts}, the reliability of perturbation theory in the vicinity of relatively weak transitions breaks down. Thus, one cannot ascertain the parameter-space boundary between a smooth crossover and first order EWPT perturbatively. Within the SM, lattice computations were decisive in revealing the onset of a crossover transition at the \lq\lq critical Higgs mass\rq\rq\, in the range of $\sim 70-80$ GeV \cite{Kajantie:1996mn, Csikor:1998eu}. A key goal of the present study is to determine this parameter space boundary for the xSM. At the same time, we assess the degree to which state-of-the-art perturbative computations -- using thermal effective theory as described below -- provide a realistic guide to the xSM EWPT thermodynamics. 

    Doing so for both the xSM and other extended scalar sector scenarios is important for both theory and experiment. In general, one would like to know: Can a given model admit a strong FOEWPT? If so, for what choices of parameters? And how can a combination of collider probes and gravitational wave (GW) searches test this possibility? These questions have recently been addressed for the real triplet scalar extension (\lq\lq $\Sigma$SM\rq\rq\, ) using a combination of lattice and thermal effective field theory methods \cite{Niemi:2018asa, Niemi:2020hto, Friedrich:2022cak}. The result is a determination of the relevant phase diagram, including a delineation of the first order phase transition crossover boundary, identification of regions of two-step EWPT, and mapping between the phase diagram and relevant experimental collider and GW sensitivities. The present study provides the foundation for a similar exploration of the xSM. 
	
	In order to provide a robust determination of the FOEWPT-viable xSM parameter space, we combine lattice simulations with perturbative computations performed using the dimensionally reduced, three-dimensional (3D) field theory at non-zero temperature (DR3dEFT). The latter allows one to obtain a comprehensive survey of the parameter space by scanning over the xSM parameters. We employ the state-of-the-art DR3dEFT at $\mathcal{O}(g^4)$, where $g$ is the generic gauge coupling and where we draw upon the necessary two-loop computations \cite{Schicho:2021gca, Niemi:2021qvp}. We carry out lattice simulations and selected benchmark points in order to determine the crossover-FOEWPT boundary and to assess the reliability of the perturbative DR3dEFT computations. We select these benchmark points based on earlier results from one-loop perturbation theory that have important phenomenological implications \cite{Profumo:2007wc, Espinosa:2011ax, Profumo:2014opa, Curtin:2014jma, Chen:2017qcz, Kurup:2017dzf, Kozaczuk:2019pet, Papaefstathiou:2020iag, Carena:2022yvx, Zhang:2023jvh}. 

	We consider two regimes based on the masses $M_{1,2}$ of xSM neutral mass eigenstates, $h_{1,2}$: (a) a heavy singlet-like scalar $h_2$ with $M_2> 2 M_1$, with $h_1$ being the SM-like Higgs boson having $M_1=125$ GeV; (b) a light $h_2$ with $M_1> 2 M_2$. Region (a) is relevant for resonant di-Higgs production in proton-proton collisions, while region (b) allows for exotic decays of the SM-like Higgs boson. For both regions, singlet-doublet mixing -- characterized by a mixing angle $\theta$ -- yields distinct signatures in precision Higgs studies, relevant for future lepton collider Higgs factories. Our key findings are: 
	\begin{itemize}
		\item[(1)] Region (a): For fixed $M_2$ and singlet Higgs portal coupling $a_2$, we find that there exists a minimum value of $|\sin\theta|$ for which a first order transition occurs, whereas perturbative computations always indicate the existence of a first order transition, even for vanishing mixing angle. 
  
		\item[(2)] Region (a): For values of $|\sin\theta|$ consistent with a FOEWPT, two-loop DR3dEFT perturbative results for the discontinuity in the Higgs scalar condensate $\langle \phi^\dag \phi\rangle$ closely track the lattice results.
  
		\item[(3)] Region (b): Perturbative computations imply that, for a given value of $M_2$, the nucleation rate is decisive for setting the minimum magnitude of the doublet-singlet cross-quartic coupling $a_2$ consistent with a FOEWPT that completes. Our lattice results indicate that, depending on the values of the other parameters in the scalar potential, thermodynamics can also play an important role in determining the minimum value of $|a_2|$.
	\end{itemize}
	
	These results have important consequences for collider phenomenology. 
	\begin{itemize}
		\item Point (1) implies that the xSM FOEWPT-viable parameter space is less extensive than one would conclude from purely perturbative studies. Thus, prospective precision Higgs and resonant di-Higgs production have a relatively greater ability to test the xSM EWPT than previously thought. 
		\item Point (2) indicates that in the regions of FOEWPT, the two-loop DR3dEFT computations provide a reliable guide to the strength of the transition. This strength is important for both the signal magnitude of potential gravitational waves result from the FOEWPT as well as for the preservation of a baryon asymmetry during electroweak baryogenesis.
		\item From point (3) we conclude that the existence and parameter space location of a minimum exotic Higgs decay branching ratio is governed by both the nucleation rate and the xSM thermodynamics. With our results for the lattice thermodynamics now in hand,  future refinements of nucleation rate computations will yield the most robust target for exotic Higgs decay studies.
	\end{itemize}
	
	The remainder of the paper provides detailed discussion of the computations and analysis leading to these conclusions. Along the way, we endeavor to provide a thorough presentation of the lattice simulations, coupled with technical material in the Appendices. Sections  \ref{sec:model} and \ref{sec:EFT} review the xSM and its DRT3dEFT formulation. Sections \ref{sec:lattice-theory}, \ref{sec:latt_gen}, and \ref{sec:find-Tc} give the corresponding lattice formulation, general features of the lattice simulations, and the procedure for obtaining the critical temperature, respectively. We define the quantities that characterize the strength of the transition in Section \ref{sec:strength}. Section \ref{sec:spacing} presents the assessment of the lattice spacing and volume dependence.  Result for regions (a) and (b) are given in Sections \ref{sec:results_heavy} and \ref{sec:small-mass}, respectively. We discuss the phenomenological implications of this work in Section \ref{sec:pheno} and summarize in Section \ref{sec:conclude}. 
    Readers who are primarily interested in model-phenomenological results may want to skip sections \ref{sec:EFT} through \ref{sec:small-mass} on the first read-through.
    There are two appendices, detailing the matching of 3D lattice and continuum theories and documenting our reweighting scheme for analyzing simulation results.

	\section{Model setup}
	\label{sec:model}
	
	The singlet-extended SM is defined by the scalar potential 
	\begin{align}
	\label{eq:scalar-pot}
	V(\phi, S) =& m^2_\phi \he\phi\phi + \lambda (\he\phi\phi)^2 + b_1 S + \frac12 m^2_S S^2 \nonumber \\
	& + \frac13 b_3 S^3 + \frac14 b_4 S^4  + \frac12 a_1 S \he\phi\phi + \frac12 a_2 S^2 \he\phi\phi
	\end{align}
	where $\phi$ is the electroweak (EW) Higgs doublet and $S$ is a real singlet scalar field. Setting $b_1, b_3, a_1 = 0$ makes the theory $Z_2$ symmetric under $S \rightarrow -S$; in this paper we focus on the more general potential where this symmetry is absent.
    
	The parameters are fixed in standard fashion, at zero temperature, by fixing a $R_\xi$ gauge  and shifting
	\begin{align}
	\phi \rightarrow \phi + \frac{1}{\sqrt{2}} \begin{pmatrix} 0 \\ v_0 \end{pmatrix},
	\end{align}
	requiring that the potential has a minimum at $v_0 > 0$ (the EW vacuum) and matching to experimental input via perturbation theory around this vacuum. We also use the freedom to shift $S$ at will to remove the vacuum expectation value (VEV) of $S$ in the EW minimum: this fixes $b_1$. Note that some studies eliminate $b_1$ in favor of the singlet VEV \cite{Profumo:2007wc, Profumo:2014opa, Zhang:2023jvh}. The model predicts two electrically-neutral scalars $h_1, h_2$ that, in the absence of aforementioned $Z_2$ symmetry, are linear combinations of the singlet and the Standard Model Higgs field $h$:
	\begin{align}
	\begin{pmatrix} 
	h_1 \\ h_2
	\end{pmatrix}
	= 
	\begin{pmatrix} 
	\cos\theta & -\sin\theta \\
	\sin\theta & \cos\theta
	\end{pmatrix}
	\begin{pmatrix} 
	h \\ S
	\end{pmatrix},
	\end{align}
	where the mixing angle $\theta$ is chosen to diagonalize the neutral scalar mass-squared matrix in the $(h_1,h_2)$ basis with $h_1$ being the SM-like Higgs boson.

	Our EW sector inputs are the $W$ and $Z$ boson pole masses, that of the top quark, the SM Higgs pole mass $M_1 = 125.10$ GeV, and the Fermi constant $G_\mu$. We neglect Yukawa interactions of all fermions lighter than the top. For BSM parameters we input the pole mass $m_2$ of the new scalar $h_2$, and $\sin\theta$. This leaves the couplings $b_3, b_4, a_2$ which we treat as free parameters (in the \MSbar scheme). Expressing parameters in the potential in terms of the inputs gives
	\begin{align}
	\label{eq:params-tree1}
	m^2_\phi &= -\frac14 \left( m^2_1 + m^2_2 - (m^2_2 - m^2_1)\cos2\theta \right)  \\
	m^2_S &= \frac12 \left( m^2_1 + m^2_2 + (m^2_2 - m^2_1)\cos2\theta - a_2 v_0^2 \right) \\
	b_1 &= -\frac14 v_0 (m^2_2 - m^2_1) \sin2\theta \\
	a_1 &= v_0^{-1} (m^2_2 - m^2_1) \sin2\theta \\
	\label{eq:params-tree2}
	\lambda &= \frac14 v_0^{-2} \left( m^2_1 + m^2_2 - (m^2_2 - m^2_2)\cos2\theta \right).
	\end{align}
	Here $m_1, m_2$ are the eigenvalues of the mass matrix. At tree level they match the pole masses, however there are corrections at loop level and importantly, at one-loop the corrections are of same parametric order as two-loop corrections to thermal masses \cite{Niemi:2021qvp}.
    In this parametrization, the $Z_2$ symmetric limit corresponds to $b_3 \rightarrow 0, \theta \rightarrow 0$.
	
	The accuracy of our lattice simulations will ultimately be constrained by the perturbative mapping between the lattice action and $T=0$ input parameters; hence we include one-loop corrections to Eqs.~(\ref{eq:params-tree1}) through (\ref{eq:params-tree2}), and to other parameters in the EW sector. We perform this in the \MSbar scheme following ref.~\cite{Niemi:2021qvp}; details of the procedure are given in the Appendix of that paper. Specificially, we fix the values of the $T=0$ parameters using Eqs.~(A27) through (A35) in \cite{Niemi:2021qvp}.

	At the perturbative level, the EW transition  occurs when the thermally-corrected effective potential shifts its global minimum from $v_0 = 0$ to $v_0 > 0$.\footnote{Note that we follow the Ehrenfest classification of phase transitions according to the lowest order derivative of the free energy that displays a discontinuity. In a crossover transition, no discontinuity occurs. If first order, some supercooling is expected before the transition actually takes place.}
    This activates the electroweak Higgs mechanism and generates tree-level masses for gauge fields. The term ``spontaneous gauge symmetry breaking'' is often used to describe this phenomenon; however we deliberately avoid this terminology as taking the concept too literally obfuscates the true, gauge invariant physics of the EWPT \cite{Kajantie:1996mn}. Indeed, physical states, including the vacuum state, are always gauge invariant, and the apparent breakdown of a global $\gr{SU(2)}$ in the $v_0 \neq 0$ minimum is an artifact of gauge fixing which is always necessary in perturbation theory \cite{Elitzur:1975im, Fradkin:1978dv}. These details become important in the nonperturbative lattice approach where gauge invariance is manifest. Gauge fixing is only needed in this and the next section for perturbatively relating the $T=0$ theory to a lattice theory.

    The main challenge in describing the finite temperature EWPT is that perturbation theory for bosons converges slowly in the infrared (IR). Specifically, at high temperature, the expansion parameter is of form $g^2 T / m$ \cite{Linde:1980ts}, where $g^2$ denotes a generic quartic coupling and $m \ll T$ is the mass of a relevant bosonic excitation. In the high-$T$ phase where the Higgs mechanism is lifted, gauge bosons are perturbatively massless and perturbative expansions are ill-behaved. This is the Linde problem \cite{Linde:1980ts} and means that any perturbative description of the EWPT is necessary incomplete. At the same time, the EWPT typically occurs at a temperature where the SM-like Higgs excitation is light compared to the temperature, which reduces reliability of perturbative predictions already in the scalar sector. The combined effect of these two problems is that perturbation theory fails to describe weak transitions (and crossovers) in which nonperturbative effects in the gauge sector are important \cite{Kajantie:1995kf}, and in other cases it is typically necessary to extend perturbative calculations beyond one-loop if accurate predictions of thermal quantities are needed \cite{Kainulainen:2019kyp, Croon:2020cgk, Gould:2022ran}.
 
    Our goal in this paper is to include the problematic IR physics by means of nonperturbative lattice simulations, and thus provide a reliable description of the EWPT in the singlet-extended model.
    With the Higgs mass $M_1$ fixed to its SM value, one expects the EWPT strength to be predominantly controlled by parameters $a_2$ and $\sin\theta$ that directly couple the singlet and the $\phi$ doublet. In contrast, $b_3$ and $b_4$ contribute only through loops and are expected to have a milder effect \cite{Profumo:2007wc, Kurup:2017dzf}. We will mostly focus on $a_2$ and $\sin\theta$.
    
    In addition to EWPT in the Higgs direction of the potential, the vacuum structure may be nontrivial also in the singlet direction. In the general, non-$Z_2$ symmetric model this is generally the case, as loop corrections shift the singlet VEV away from the origin. A similar situation can arise in the $Z_2$ symmetric limit if the theory undergoes spontaneous breakdown of the discrete $Z_2$ symmetry at temperatures above the EWPT temperature. This is the two-step phase transition scenario \cite{Profumo:2007wc, Espinosa:2011ax, Curtin:2014jma, Ramsey-Musolf:2019lsf}, and analogous multi-step transitions can occur also in the more general model but without spontaneous $Z_2$ breaking \cite{Gould:2021dzl}. Transitions in the pure singlet direction will not be discussed further in this work as our focus is on the actual EWPT.

	\section{High-T effective theory}
	\label{sec:EFT}
 
	Some approximations are required before the finite-temperature theory can be studied nonperturbatively on the lattice. 
    The most fundamental issue to overcome is that formulating a chiral gauge theory such as the SM on the lattice is an unsolved problem (see \cite{Kaplan:2023pxd} for a recent development).\footnote{Not to be confused with technical challenges associated with implementing chiral symmetry and its breaking on the lattice in non-chiral gauge theories such as QCD.}
    The second issue is that controlling the continuum limit while simultaneously preserving connection to known EW physics is tedious when the theory contains many parameters, each requiring renormalization. The final hurdle is that a finite-temperature system admits a natural hierarchy of scales, {\it e.g.} scales $gT$ and $g^2T$ associated with screening of electric and magnetic fields respectively and the scale $\pi T$ of short-distance thermal fluctuations. Fitting these on a lattice simultaneously necessitates numerically demanding simulations. 

    Despite these complications, simulations and continuum extrapolations have been successfully carried out for the bosonic $\SU2$ + Higgs theory at finite temperature \cite{Csikor:1998eu,Aoki:1999fi}. Still, for the singlet extension this approach seems unpractical due to the large number of parameters in the scalar potential, and would require that we neglect all effects of fermions on the EWPT; likely not a good approximation given that the thermal Higgs mass obtains its largest SM contributions from the top quark \cite{Kajantie:1995dw}.

    A way of bypassing the aforementioned issues is provided by the effective field theory (EFT) approach to high-$T$ field theory \cite{Ginsparg:1980ef,Appelquist:1981vg,Braaten:1995cm,Kajantie:1995dw}. In this method, one perturbatively maps the 4D finite-$T$ theory to a simpler 3D theory with similar field content (\lq\lq\,dimensional reduction\rq\rq\,), apart from fermions whose contributions are included in \lq\lq matching relations\rq\rq\, between the 3D and 4D parameters. The resulting bosonic 3D EFT can then be simulated on the lattice, avoiding all of the issues highlighted above \cite{Farakos:1994xh}. This 3D approach is common for EWPT studies both in lattice \cite{Kajantie:1995kf,Kajantie:1996qd,Laine:1998qk,Laine:2000rm,Kainulainen:2019kyp,Niemi:2020hto} and perturbative contexts \cite{Arnold:1992rz, Farakos:1994kx, Bodeker:1996pc, Niemi:2021qvp, Gould:2021oba, Ekstedt:2022zro, Gould:2023ovu}, and it is the approach we follow in this paper.
    The downside is that the quality of 4D $\rightarrow$ 3D mapping becomes a limiting factor for the full analysis. This is usually not an issue as long as the zero-temperature theory is sufficiently weakly coupled. For instance the 4D and 3D simulations of the $\SU2$ + Higgs case agree within error bars \cite{Laine:1999rv}. Systematic errors are in much better control in the 3D analysis because the effective 3D theory is super-renormalizable, making it possible to find analytical relations between continuum and lattice actions that become exact in the continuum limit \cite{Laine:1995np, Laine:1997dy}. This generalizes directly to the singlet extension and is the final step in our pipeline relating the \lq\lq physical\rq\rq\, input parameters, described in section~\ref{sec:model}, to a practical lattice formulation.

	For our model the high-$T$ EFT is a super-renormalizable $\SU2 \times \grU1$ + Higgs + singlet theory in 3D with temperature dependent parameters. The action reads
	\begin{widetext}
		\begin{align}
		\label{eq:3D-EFT}
		S_\text{3D} &= \frac{1}{T} \int d^3x \; \Big\{ \frac14 F^a_{ij} F^a_{ij} + \frac14 B_{ij} B_{ij} + |D_i \phi|^2 + \frac12 (\partial_i S)^2 + V_{\text{3D}}(\phi, S) \Big\} \\
		V_{\text{3D}}(\phi, S) &= \bar{m}_\phi^2 \he\phi\phi + \bar{\lambda} (\he\phi\phi)^2 + \bar{b}_1 S + \frac12 \bar{m}_S^2 S^2 + \frac13 \bar{b}_3 S^3 + \frac14 \bar{b}_4 S^4 + \frac12 \bar{a}_1 S \he\phi\phi + \frac12 \bar{a}_2 S^2 \he\phi\phi.
		\end{align}
	\end{widetext}
	In the above, $F_{ij}$ and $B_{ij}$ are $\SU2$ and $\grU1$ field strengths in 3D for their respective gauge fields $A_i$ and $B_i$. The Higgs covariant derivative is $D_i \phi = (\partial_i + i \bar{g} A_i + \frac12 i \bar{g}' B_i)\phi$. The prefactor $T^{-1}$, arising from trivial integration over the imaginary time, is often removed by a rescaling of fields and couplings in the EFT, as done for example in refs.~\cite{Kajantie:1995kf, Kainulainen:2019kyp, Niemi:2020hto, Niemi:2021qvp} (but not in \cite{Laine:2000rm}). We have chosen to not perform this rescaling in order to keep the connection to 4D quantities and units more transparent.
	The 4D $\rightarrow$ 3D relation for the parameters has been derived in \cite{Niemi:2021qvp} and can be read from  Eqs.~(30)-(41) and (51)-(55) therein.\footnote{In addition to keeping the $1/T$ factor explicit, we have simplified the notation from that of Ref.~\cite{Niemi:2021qvp} by dropping redundant subscripts from the 3D parameters. For instance, our $\bar{a}_2$ corresponds to $\bar{a}_{2,3}/T$ in \cite{Niemi:2021qvp}.}
	The matching relations from \cite{Niemi:2021qvp} include thermal mass corrections at $\mathcal{O}(g^4)$, or two-loop, and corrections to coupling constants at the same parametric order.

	The 3D approximation is formally valid for energy scales $\ll \pi T$, {\it i.e.} when the high-$T$ expansion of loop integrals is valid.
    The dominant error arises from neglect of higher-dimensional operators in the 3D action: for example, the 4D $\rightarrow$ 3D matching generates dimension five and six operators such as $S^5$ and $(\he\phi\phi)^3$. These contribute only at higher orders of high-$T$ perturbation theory ({\it i.e}. we expand in $m/T$) \cite{Kajantie:1995dw,Bodeker:1996pc,Niemi:2021qvp} and are neglected in our numerical analysis (see \ref{sec:lattice-theory} for more discussion on this point). In our model we treat the $T=0$ BSM scalar mass $M_2$ as an input, and EFT will likely fail if the ratio $M_2/T$ grows large. This is a qualitative statement and cannot directly be used to set a strict upper bound on allowed $M_2$ because the relevant finite-$T$ mass is different from $M_2$, and the way $M_2$ affects EWPT quantities is not easily tractable.

    A quantitative error estimate for the truncated EFT (\ref{eq:3D-EFT}) can be obtained by explicitly including operators at dimensions five and six and perturbatively calculating the shift induced in the effective Higgs potential\footnote{Specifically, location of its minimum after the EWPT. This is not a direct physical quantity but is still a useful indicator of EWPT strength in the perturbative context.}
    Doing this in the SM case suggests percent-level accuracy for the 4D $\rightarrow$ 3D mapping \cite{Kajantie:1995dw}. This was generalized to the singlet extension in \cite{Niemi:2021qvp} and the error was found to remain small, much less than $5\%$ for most benchmarks in the reference. Hence the 3D EFT should provide an excellent starting point also for precision lattice studies at least for $M_2 \lesssim$ $450$ GeV which was the largest $M_2$ studied in \cite{Niemi:2021qvp}.

	\section{Lattice formulation}
	\label{sec:lattice-theory}
	
	It is straightforward to write down a lattice action that reduces to the EFT (\ref{eq:3D-EFT}) in the naive continuum limit $a\rightarrow 0$, $a$ being the lattice spacing. For completeness, we include also the $\grU1$ hypercharge interactions that are often neglected in the EWPT simulations (for lattice studies of the SM transition including the $\grU1$ field, see \cite{Kajantie:1996qd,Kajantie:1998rz,Annala:2023jvr}). Our lattice action reads
	\begin{widetext}
		\begin{align}
		\label{eq:lattice-act}
		S_L &= \beta \sum_{x, i<j} \left[ 1 - \frac{1}{2} \RE \Tr P_{ij}(x) \right] + \beta' \sum_{x, i<j} \Big[ 1 - \RE p_{ij}^{r}(x) \Big] \nn 
		& \; + 2 aT^{-1} \sum_{x,i} \Big[ \he\phi(x)\phi(x) - \he\phi(x) U_i(x) u_i(x) \phi(x+i) \Big]  + aT^{-1} \sum_{x,i}\Big[ S(x)^2 - S(x)S(x+i) \Big] \nn 
		&\; + a^3 T^{-1} \sum_x \Big[ m_{\phi,L}^2 \he\phi\phi + \bar{\lambda} (\he\phi\phi)^2 + b_{1,L} S + \frac12 m_{S,L}^2 S^2 + \frac13 \bar{b}_3 S^3 + \frac14 \bar{b}_4 S^4 + \frac12 \bar{a}_1 S \he\phi\phi + \frac12 \bar{a}_2 S^2 \he\phi\phi \Big].
		\end{align}
	\end{widetext}
	Here $U_i$ and $u_i$ are $\SU2$ and $\grU1$ link variables respectively. At small $a$ their relation to the continuum gauge fields is
	\begin{align}
	U_i(x) = e^{i a \bar{g} A_i(x)}, \quad u_i(x) = e^{\frac12 i a \bar{g}' B_i(x)}.
	\end{align}
	$P_{ij}$ and $p_{ij}$ are elementary plaquettes constructed from the links,
	\begin{align}
	P_{ij}(x) = U_i(x) U_j(x+i) \he U_i(x+j) \he U_j(x)
	\end{align}
	and similarly for the $\grU1$ plaquette $p_{ij}$. Couplings appearing in the gauge part are
	\begin{align}
	\beta &= \frac{4}{a T \bar{g}^2}, \quad\quad
	\beta' = \frac{4}{a T \bar{g}'^2 r^2}
	\end{align}
	and $r \neq 0$ is an integer that labels irreducible representations of the $\grU1$ group. We use standard periodic boundary conditions to preserve translational invariance.
	
	Parameters appearing in the lattice action are the unrenormalized (bare) ones, and it remains to match these to the \MSbar renormalized parameters of the EFT (\ref{eq:3D-EFT}). Because of super-renormalizability, only $m_{\phi,L}^2, m_{S,L}^2$ and $b_{1,L}$ contain divergences, and the other parameters are renormalization group (RG) invariant up to corrections that vanish in the $a\rightarrow 0$ limit. Following Ref.~\cite{Laine:1995np}, we may analytically relate the divergent parameters to those appearing in the continuum theory by calculating the 3D energy density around a generic $(\phi, S)$ background, ({\it i.e.}, the effective potential) and equating the lattice and continuum results. A two-loop calculation suffices to remove all divergences. Moreover, since the two schemes can differ only in the UV region, nonperturbative IR effects do not spoil the lattice-continuum matching.

	Appendix ~\ref{sec:lattice-counterterms} describes the lattice-continuum matching calculation in more detail. Eqs.~(\ref{eq:lattice-ct-start})-(\ref{eq:lattice-ct-end}) give the resulting lattice-continuum relations for $m_{\phi,L}^2, m_{S,L}^2, b_{1,L}$, which is exact up to $\mathcal{O}(a)$ corrections. Remaining cutoff dependence can be eliminated by extrapolating simulation results to the continuum, see section \ref{sec:spacing}. 

    As indicated above, this analytical mapping between continuum and lattice actions allows us to fix renormalized parameters on the continuum side so that we reproduce perturbative $T=0$ physics of the EW sector, perform the finite-$T$ reduction from 4D to 3D, and finally compute the corresponding lattice parameters at given value of the lattice spacing. This avoids the need for nonperturbative renormalization.

    We note in passing that it is, in principle, possible to extend the lattice action with operators of higher dimensionality, for instance the aforementioned $(\he\phi\phi)^3$. This could be useful if the 3D EFT (\ref{eq:3D-EFT}), truncated at operator dimension four, is deemed too inaccurate in some region of the parameter space. However, operators of dimension six are not super-renormalizable in 3D, and analytical control over the $a \rightarrow 0$ limit is then lost. In our view, these complications makes simulations with higher-order operators highly impractical.

	At finite cutoff the dynamics depends on the representation chosen for $\grU1$, in particular the lattice theory admits topological monopole configurations \cite{Polyakov:1975rs} that however vanish in the $a \rightarrow 0$ limit. For simplicity we perform all of our simulations at $r = 1$. As discussed in \cite{Kajantie:1996qd}, this choice minimizes correction terms in the lattice-continuum relations leading to smaller $\mathcal{O}(a)$ effects related to $\bar{g}'$. A potential complication is that the ratio $\beta' / \beta$ is rather large for realistic values of $\grU1$ and $\SU2$ couplings, necessitating a large number of lattice sites to fully capture the IR behavior of the hypercharge field. This situation is not too worrying, however, considering that the effect of $\grU1$ on the EWPT is small in the first place \cite{Kajantie:1996qd}, and our results below indicate a clean infinite-volume limit.

	\section{Lattice simulations: Generalities}
    \label{sec:latt_gen}
	
    The EW theory (with or without the singlet) does not admit a gauge-invariant order parameter, nor a symmetry, that could distinguish phases of active and inactive Higgs mechanism (\lq\lq broken\lq\lq\, and \lq\lq symmetric\lq\lq\, phases, or Higgsed and non-Higgsed phases). This is a generic feature of gauge-Higgs theories with fundamental representation matter \cite{Osterwalder:1977pc, Fradkin:1978dv} and means that there really is just one thermodynamical phase, one that is strongly coupled at high temperatures due to the Linde problem, and weakly coupled at low temperatures where tree-level masses produced by the Higgs mechanism yield an IR cutoff in the gauge sector. We say that the phase diagram is continuously connected. 
    
    Keeping the above in mind, in the following we use the common labels \lq\lq symmetric\lq\lq\, and \lq\lq broken\lq\lq\, phases to describe high-$T$ and low-$T$ regimes of the theory respectively. The EWPT is the transition between these two regimes as the temperature changes. This transition can occur smoothly as in the SM (crossover, {\it i.e.} not a phase transition at all), or be abrupt and discontinuous as in a first-order phase transition. At the boundary where first-order behavior turns into a crossover there is a second order phase transition \cite{Kajantie:1996mn, Rummukainen:1998as}. The situation is analogous to the liquid-gas transition in classical fluids such as water, also a smooth crossover at high temperature and pressure.

    Nonperturbative studies of the EWPT are based on expectation values of local gauge-invariant operators, or condensates, and their distributions in the canonical ensemble. In our singlet extension the simplest operators of interest are $\he\phi\phi$, $S$ and $S^2$.
    The Higgs condensate $\langle \he\phi\phi \rangle$ is discontinuous at a first-order EWPT \cite{Farakos:1994xh, Kajantie:1996mn} and acts as an effective order parameter for the transition. In the $Z_2$ symmetric limit of the theory, a nonzero value of $\langle S \rangle$ signals spontaneous breakdown of the discrete symmetry, but does not cause Higgs mechanism. In the more general model in which the $Z_2$ symmetry is explicitly broken by cubic interactions in the scalar potential, $\langle S \rangle$ is generally non-zero at all temperatures and the phase diagram is expected to again be continuously connected.

	A Monte Carlo simulation aims to generate field configurations in the canonical ensemble, {\it i.e.} the probability distribution of a generic lattice field configuration $\{\varphi\}$ is $p(\{\varphi\}) \propto \exp[-S_L(\{\varphi\})]$. Condensates such as $\langle \he\phi\phi \rangle$ are estimated by measuring $\he\phi(x)\phi(x)$ locally and averaging over the lattice to reduce noise. For a sufficiently large sample $\{ \varphi_i \}$ of canonically-distributed field configurations, the mean of individual measurements yields a good estimate for the expectation value $\langle \he\phi\phi \rangle$.
	
	Standard sampling algorithms for generating field configurations with the distribution $p(\{\varphi\}) \propto \exp[-S_L(\{\varphi\})]$ proceed through local deformations of the fields with a probabilistic accept/reject step to preserve the ensemble. As such, successive field configurations are not statistically independent, and the field update algorithm should be designed to efficiently reduce autocorrelations in observables of interest.
	Our field update algorithm combines standard $\gr{SU(2)}$ heatbath \cite{Kennedy:1985nu} for the gauge links and the Higgs overrelaxation method of Kajantie {\it et al.} for the scalars \cite{Kajantie:1995kf}, modified to include the singlet field.  Compared to simpler Metropolis-like algorithms, the overrelaxation method more effectively evolves the radial Higgs field mode \cite{Kajantie:1995kf}. We observe similar improvement also for the singlet (see appendix~D in \cite{Gould:2021dzl} for a related comparison). Our full update sweep consists of one heatbath update of gauge links and four scalar overrelaxation steps, followed by a Metropolis update on the scalars to guarantee ergodicity.

	\begin{figure}[!htb]
		\includegraphics[width=\columnwidth]{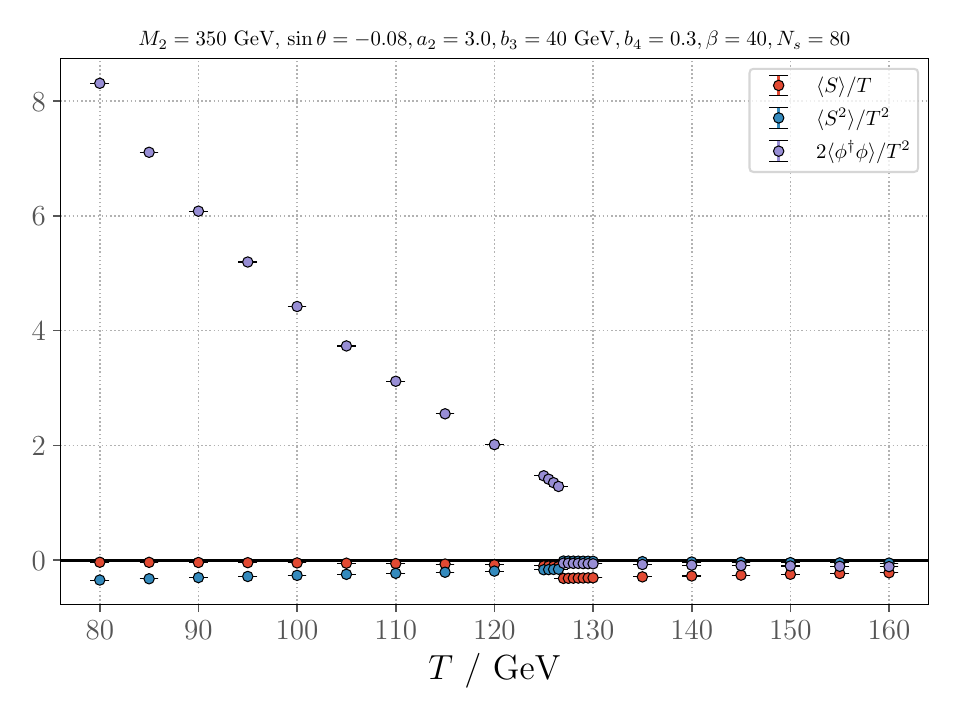}
		\caption{Temperature dependence of scalar condensates in the benchmark point defined in Eq.~(\ref{eq:BM3}) below, converted to \MSbar scheme (scale $\bar{\mu} = T$). 
        The data is taken on a $\beta = 40, N_s = 80$ lattice. Each point is from a separate short simulation consisting of $4000$ - $10000$ measurements. Negative values of $\langle \he\phi\phi \rangle$ and $\langle S^2\rangle$ are due to additive renormalization, see Eq.~(\ref{eq:condensate-renormalization}).}
		\label{fig:BM3_condensates}
	\end{figure}

	To find the phase transition point we perform simulations at different temperatures, measuring condensates of $\he\phi\phi$ and of other operators appearing in the action and tracking their evolution with the temperature. Fig.~\ref{fig:BM3_condensates} illustrates this temperature evolution for scalar condensates of lowest dimensionality in one benchmark scenario. At high temperatures, the Higgs condensate $\langle \he\phi\phi \rangle$ is approximately constant in units of $T$ but undergoes a sudden change to a larger value at $T \approx 127$ GeV. In this lower-$T$ regime, the conventional Higgs mechanism becomes effective.
    The $\langle \he\phi\phi \rangle$ discontinuity seen in Fig.~\ref{fig:BM3_condensates} suggests a first-order transition, and means physically that the low- and high-$T$ phases can coexist, separated by a free energy barrier.\footnote{Strictly speaking the condensate can be discontinuous only in the thermodynamic limit ($V\rightarrow \infty$), while in simulations we operate on a finite-size system. At finite volume $\langle \he\phi\phi \rangle$ can take values outside the thermodynamically allowed bulk values, but this is exponentially unlikely when the volume is sufficiently large. There is no ambiguity in deducing that we have a first-order transition if the probabilistic suppression persists and increases indefinitely as the volume grows.}
    The first quantity of interest for EWPT analyses is the critical temperature $T_c$, at which both phases are thermodynamically equally favored. 
    
    The discontinuous behavior contrasts with that of a crossover transition in which the condensates interpolate smoothly and no phase boundary exists. As an aside, we point out that the condensates would remain continuous also in a second-order phase transition. The main practical difference between second order transition and crossover is that in the former the system develops an infinite correlation length ({\it i.e.} massless Higgs excitation) and exhibits critical behavior. The EWPT is known to be second order at the boundary between first-order and crossover transitions \cite{Kajantie:1996mn, Rummukainen:1998as} and is therefore a very specialized case. 

	A few remarks regarding Fig.~\ref{fig:BM3_condensates} are now in order:
	\begin{enumerate}[i.]
		\item In conventional perturbative studies based on the thermal effective potential for the scalar fields, one finds the phase transition point by following the global minimum of the potential as function of the temperature. The lattice approach replaces minimization of the potential by a Monte Carlo determination of scalar condensates from thermalized field configurations (however, see point iii. below).

		\item Condensates of composite operators are generally UV divergent and carry no direct physical meaning. As for the lattice mass parameters in Section~\ref{sec:lattice-theory}, we may relate the bare (unrenormalized) lattice condensates obtained from simulations to renormalized condensates in a given renormalization scheme, cf. \cite{Farakos:1994xh, Laine:1995np} and Appendix~\ref{sec:lattice-counterterms}. For Fig.~\ref{fig:BM3_condensates}, we have converted the nonperturbative lattice condensates to continuum \MSbar scheme (scale $\bar\mu = T$). For quadratic condensates like $\langle \he\phi\phi \rangle$ and $\langle S^2 \rangle$ the required renormalization is additive and drops out when considering condensate discontinuities across a phase transition \cite{Farakos:1994xh}. The non-composite singlet condensate $\langle S \rangle$ does not require renormalization.

		\item For each $T$, we start the simulation from a ``cold'' field configuration. For example, the fields are set to a constant value everywhere, and the field update algorithm is guaranteed to bring the system into a thermalized state after an initial thermalization period.\footnote{Naturally, measurements should be taken only after the system has fully thermalized. Different observables can have different thermalization times; here we start measurements once the volume-averaged $\he\phi\phi$ reaches a stable value. Typically this occurs well within 500 iterations of our lattice update, although thermalization can take longer near the critical temperature.}
		However, if the theory admits multiple (meta)stable phases ({\it i.e.}, distinct local minima of the free energy) at a given $T$, there is no guarantee that the simulation thermalizes to the global free energy minimum instead of a metastable branch of a long-lived but statistically disfavored phase. For instance, a simulation at $T < T_c$ may get stuck in the symmetric phase and not be able to tunnel to the more stable broken phase.
		
	\end{enumerate}
	As illustrated by point iii, it is necessary to have reliable means of comparing free energies of the two phases. We do so by studying probability distributions of the order parameter, {\it i.e.} histograms of $\he\phi\phi$.
	
	Near a first order transition the histogram has a characteristic two-peak structure as depicted in Fig.~\ref{fig:hgrams}. The peaks describe bulk fluctuations of the condensate in two minima of the free energy. At critical temperature the phases have equal free energies. One may formalize this condition by requiring that both peaks in the order parameter histogram cover equal areas at $T=T_c$ \cite{Kajantie:1995kf}.
	
	As evident from Fig.~\ref{fig:hgrams}, the probability of finding configurations where $\he\phi\phi$ lies outside its thermodynamically preferred value is exponentially suppressed. At finite volume, most configurations in the disfavored region are inhomogeneous mixed-phase configurations in which a subregion of the lattice is in one metastable phase and rest of the lattice is in the other phase (see \textit{e.g.} \cite{Moore:2000jw}). The free-energy cost of such configuration is proportional to the area of the phase interface. Thus, the suppression of disfavored configurations increases with system size.

	\section{Finding the critical temperature}
	\label{sec:find-Tc}
	
	\begin{figure}[t]
		\includegraphics[width=\columnwidth]{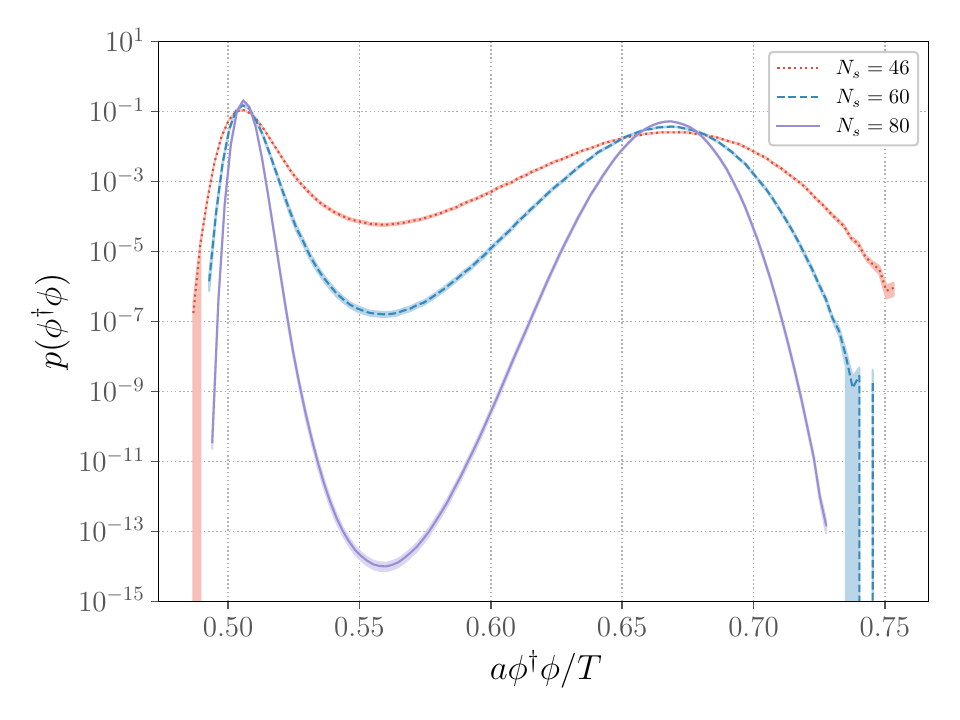}
		\caption{Normalized histograms of the quadratic Higgs condensate at different volumes ($N_s^3$ total sites), evaluated at the critical temperature for benchmark point (\ref{eq:BM3}) with $\beta = 40$ using multicanonical simulations. The peaks at small and large $a\he\phi\phi/T$ correspond to the symmetric and broken phases, respectively. Statistical errors are denoted by shaded bands.
        }  
		\label{fig:hgrams}
	\end{figure}

	In conventional simulations that sample field configurations in the canonical ensemble, the probability of any configuration that does not correspond to a thermodynamically stable phase is exponentially small. But because the configurations are sampled through small deformations of the fields, the simulation will necessarily need to pass through disfavored configurations in order to transition from one phase to another. For large enough volumes this will never happen on practical time scales, and canonical simulations are often limited to sampling only one phase.
	
	A standard way of overcoming this difficulty is to use multicanonical simulations \cite{Berg:1992qua}. This approach consists of  modifying the ensemble by shifting $S \rightarrow S + W(\he\phi\phi)$, where $W$ is a carefully chosen weight function. Distribution of the order parameter in this modified ensemble is 
	\begin{align}
	\label{eq:p_multi}
	p_\text{multi}(\he\phi\phi) \propto p_\text{can}(\he\phi\phi) e^{-W(\he\phi\phi)}.
	\end{align}
	Here $p_\text{can}$ is the canonical probability distribution. In a multicanonical simulation, we compute $p_\text{multi}(\he\phi\phi)$ by sampling configurations in the modified ensemble.\footnote{Following \cite{Laine:1998qk}, this is achieved by performing a global accept/reject step based on change in $W$ after updating half of lattice sites. The local field updates are performed as in standard canonical simulations, and the additional multicanonical step biases the simulation to prefer configurations with small $W(\he\phi\phi)$.}
	The multicanonical distribution becomes flat if the weight is chosen according to $W(\he\phi\phi) = \ln p_\text{can}(\he\phi\phi)$, and the simulation can freely probe all values of the order parameter in a random walk fashion. This bypasses the probabilistic suppression seen in Fig.~\ref{fig:hgrams}. To extract physical information about the phase transition we ultimately need the canonical histogram, obtained by inverting Eq.~(\ref{eq:p_multi}).
	
	An obvious drawback of the multicanonical method is that the optimal weight function $W(\he\phi\phi) = \ln p_\text{can}(\he\phi\phi)$ depends on $p_\text{can}(\he\phi\phi)$, the quantity we are trying to obtain in the first place. In practice we have to find a weight function that suffices to allow efficient sampling in the interesting order parameter range. We do so by an automated recursive process akin to refs.~\cite{Kajantie:1995kf,Laine:1998qk}, consisting of short simulations to estimate $p_\text{can}(\he\phi\phi)$. Once sufficient convergence for $W$ has been reached, we perform a production run with fixed weight function.
	
	We identify the critical temperature by the condition that both phases have same free energy at $T = T_c$, {\it i.e.}, peaks in the canonical histogram cover equal areas. In practice, we first scan (as in Fig.~\ref{fig:BM3_condensates}) with small-volume simulations and look for an approximate $T \approx T_c$ that produces a two-peak histogram. We then run a long multicanonical simulation at the target volume to obtain a clean two-peak histogram. Finally, we vary the temperature in order to obtain the critical histogram with equally-probable peaks.\footnote{The practical condition is as follows. We specify a ``cutpoint'' near the value of $\he\phi\phi$ where the histogram has a local minimum, and count how many measurements lie to left and right of the cutpoint. For $T=T_c$ the counts should be equal. The precise choice of the cutpoint does not matter as near the probability minimum there is only an exponentially small number of measurements.}
	
	We perform the last step without resorting to new simulations by using a standard reweighting technique \cite{Ferrenberg:1988yz}. In brief, reweighting exploits the relationship between distributions of a configuration $\{ \phi \}$ at temperatures $T$ and $T'$:
	\begin{align}
	p_{T'}(\{ \phi \}) \propto e^{-S_{T'} (\{ \phi \}) + S_{T}( \{ \phi \} ) } p_T(\{ \phi \}) \ \ \ ,
	\end{align}
    thereby allowing us to use data from a fixed-$T$ simulation to obtain expectation values and histograms also at any other $T$ that is reasonably close to the simulated temperature.  Details of our reweighting scheme are presented in Appendix~\ref{sec:reweight}.

	We also utilize an alternative for finding the $T_c$, based on mixed-phase configurations where two phases exist simultaneously on the lattice and separated by a planar interface (there will be two interfaces on a periodic lattice). At the critical temperature both phases will, on average, occupy equal fraction of the total volume, and the order parameter histogram becomes flat when restricted to mixed-phase configurations only. This method has previously been discussed in \cite{Moore:2019lua,Niemi:2020hto,Gould:2022ran} and works well with ``cylindrical'' lattices where one dimension is longer than the others; this guarantees that the interface(s) form perpendicular to the long direction. Mixed-phase configurations can be sampled by manually choosing a multicanonical weight function that traps the order parameter in a small window between its statistically favored values.

	The mixed-phase method is convenient for very strong transitions where the computation of multicanoncal weights for the standard algorithm can become costly. The downside is that one does not get full probability distributions, and separate simulations in both phases are needed to find values of condensates.
	
	\section{Strength of the transition}
	\label{sec:strength}
	
	The quantities commonly used for quantifying EWPT strength are the discontinuity of the Higgs condensate\footnote{Note that our $v$ is different from what is usually meant by the Higgs VEV in perturbative context, {\it i.e.} minimum of the effective potential in a fixed gauge. The minimum is gauge-dependent by construction and does not make a good observable, however in $R_\xi$ Landau gauge its value is known to be close to the gauge-invariant quantity defined in (\ref{eq:higgs-vev}) \cite{Niemi:2021qvp}.} 
	\begin{align}
	\label{eq:higgs-vev}
	\frac{v^2}{2} \equiv \Delta \langle \he\phi\phi \rangle,
	\end{align}
	where $\Delta(\cdots)$ means difference of low- and high-$T$ phases, and the latent heat $L$. The condensate discontinuity (\ref{eq:higgs-vev}) is both gauge- and RG-invariant (in 3D) and is correlated with the suppression of electroweak sphaleron rate in the broken phase \cite{DOnofrio:2014rug}. In contrast, the latent heat describes the amount of energy released in the transition and is more relevant for gravitational-wave studies \cite{Espinosa:2010hh}. In the minimal SM (without the $\grU1$ field) these quantities are proportional \cite{Farakos:1994xh}, but in a BSM setting their relationship is more complicated as the latent heat obtains contributions from other condensates as well. 
	
	Jump in the Higgs condensate is directly obtainable from histogram such as those in Fig.~\ref{fig:hgrams}. For extracting the latent heat we employ two methods:
	\begin{enumerate}
		\item In terms of the free energy density $f$ we have, at $T=T_c$,
		\begin{align}
		\label{eq:latent_prob}
		L = -T_c \der[\Delta f]{T} = \frac{T_c^2}{V} \der{T} \ln \frac{P_2}{P_1},
		\end{align}
		where $P_1$ and $P_2$ are probabilities for finding the system in phase 1 or 2, respectively \cite{Kajantie:1995kf}. The ratio $P_2/P_1$ is found from histograms of the order parameter $\he\phi\phi$
        as described above (at the critical temperature, $P_1 = P_2$), and the $T$-derivative is calculated by reweighting, see appendix~\ref{sec:reweight}.

		\item The above can equally be written as 
		\begin{align}
		\label{eq:latent_dSdT}
		L = \frac{T_c^2}{V} Z^{-1}\der T \int D\phi \; e^{-S_L} = - \frac{T_c^2}{V}\Delta \Big\langle \der[S_L]{T} \Big\rangle.
		\end{align}
		Here $Z$ is the partition function and $S_L$ the lattice action. We measure this expectation value from separate simulations in the two phases; this is possible once the volume is large enough that a non-multicanonical simulation never tunnels to the other phase. In practice we evaluate $dS_L/dT$ by expressing it in terms of condensates via the chain rule, see related expression (\ref{eq:action-volume-avg}) in the appendix.
		
	\end{enumerate}
	
	The latter method can be applied without knowledge of the full order-parameter probability distribution, provided that $T_c$ is found by other means, for example using the mixed-phase approach discussed at the end of section~\ref{sec:find-Tc}.

	\begin{figure*}[tb]
		\subfloat[]{\includegraphics[width=0.5\textwidth]{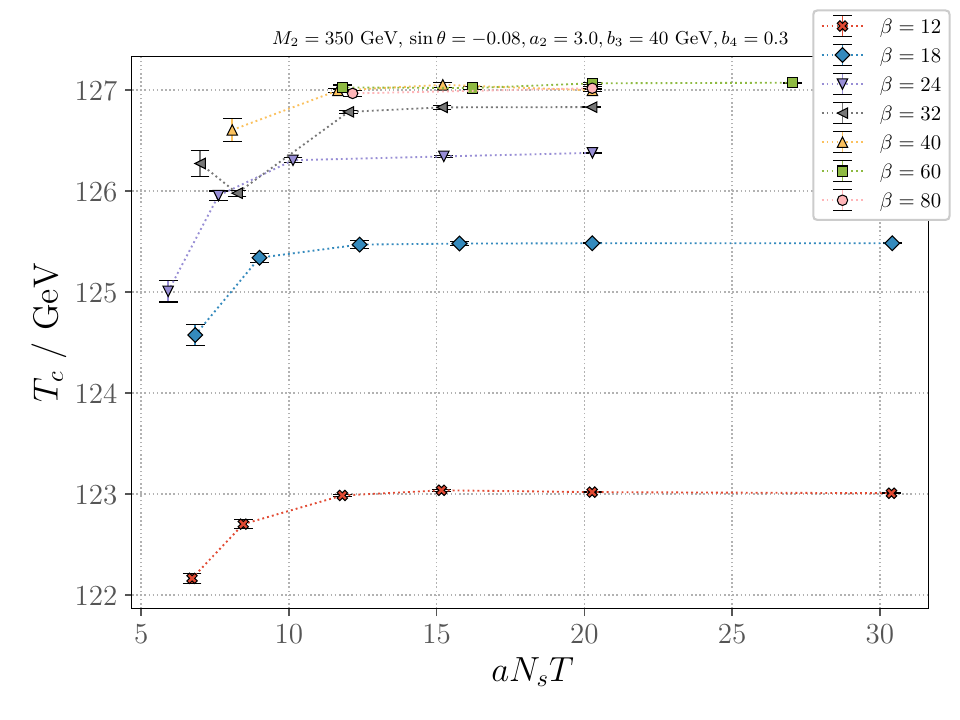}}
		\subfloat[]{\includegraphics[width=0.5\textwidth]{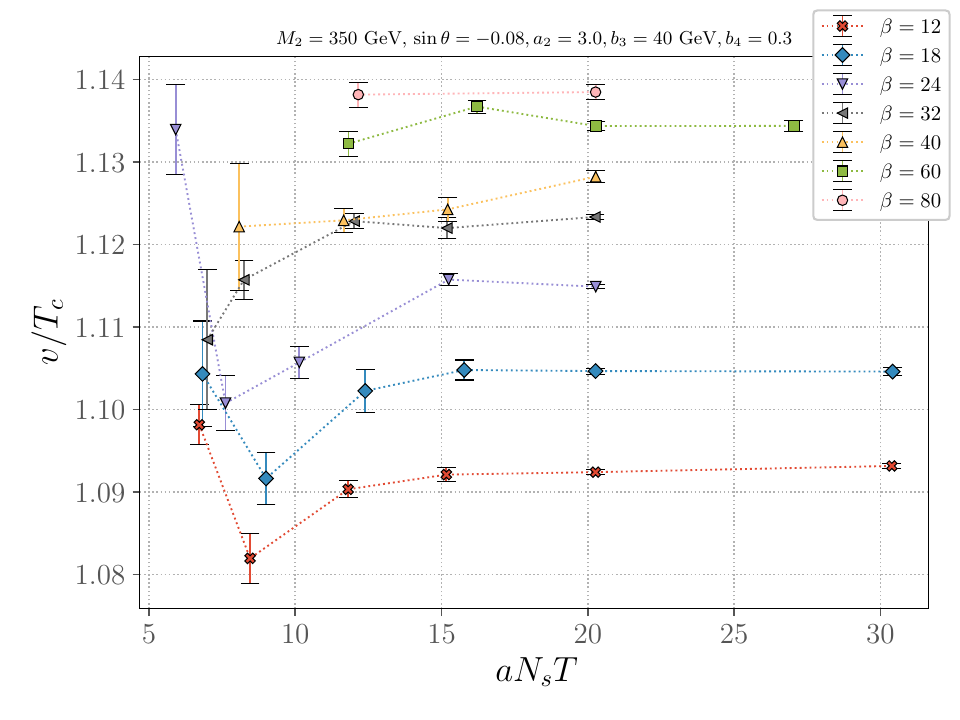}}
		\caption{Simulation results for the critical temperature (left) and the Higgs condensate discontinuity (\ref{eq:higgs-vev}) (right) in the benchmark point (\ref{eq:BM3}). Points with same $\beta$ are connected with dashed lines to guide the eye.}
		\label{fig:BM3_volume}
	\end{figure*}
 
	\section{Lattice spacing and volume dependence}
	\label{sec:spacing}
	
	\begin{figure*}[t]
		\subfloat[]{\includegraphics[width=0.5\textwidth]{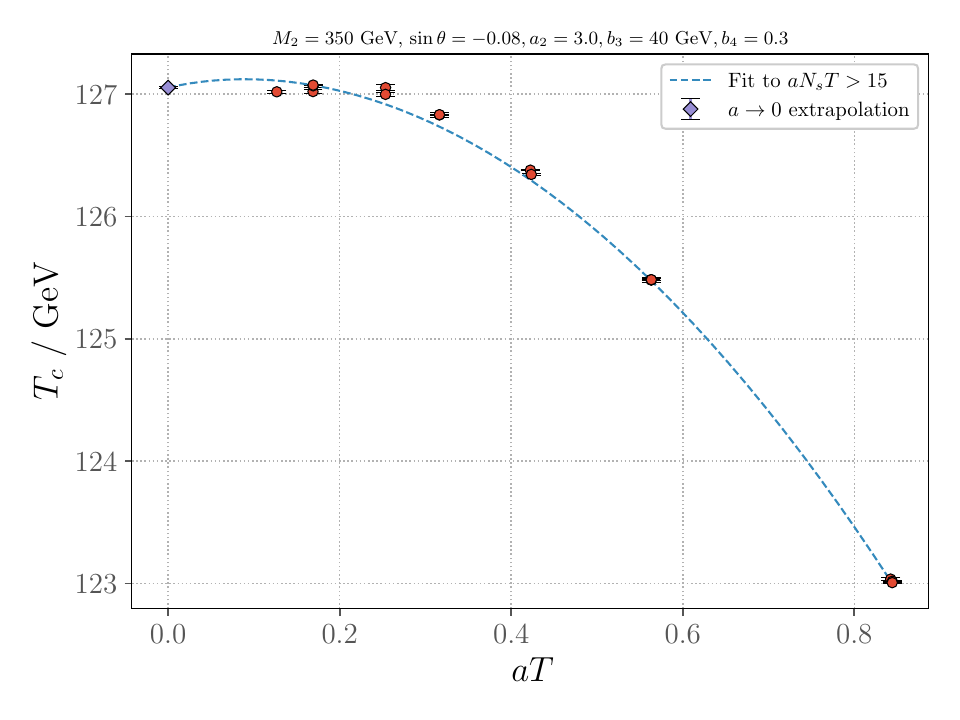}}
		\subfloat[]{\includegraphics[width=0.5\textwidth]{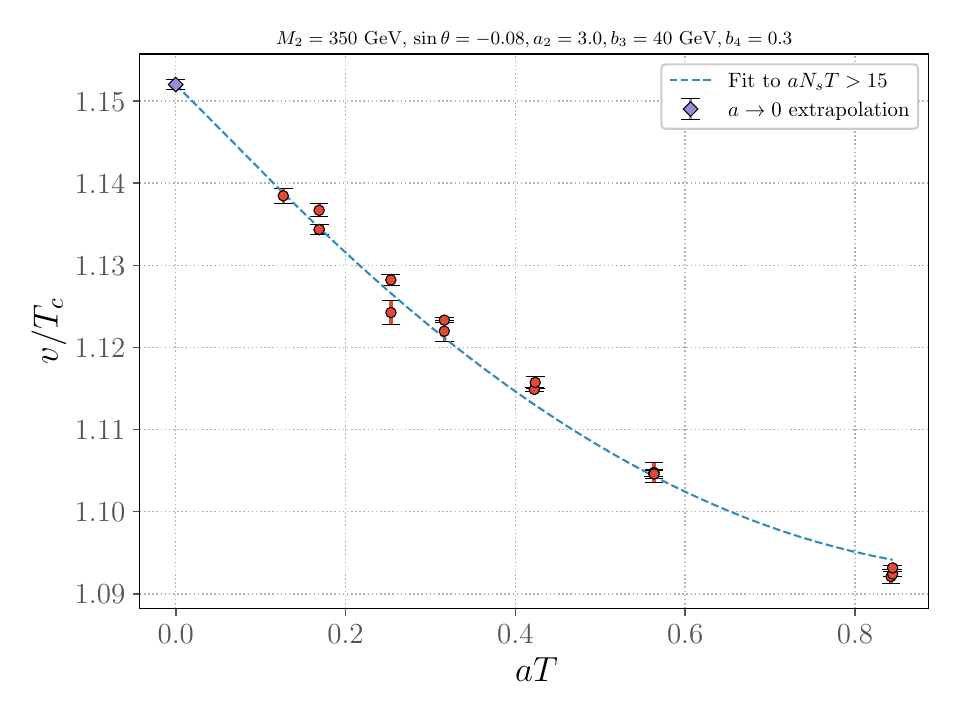}}
		\caption{Continuum extrapolations for $T_c$ and the Higgs condensate discontinuity. We have used fit functions of form $1 + aT(1 + (aN_sT)^{-1}) + (aT)^2 (1 + (aN_sT)^{-1})$ and included only the large-volume runs with $aN_sT > 15$. }
		\label{fig:BM3_spacing}
	\end{figure*}

	\begin{figure*}[!htb]
		\subfloat[Method 1: Using Eq.~(\ref{eq:latent_prob}).]{\includegraphics[width=0.5\textwidth]{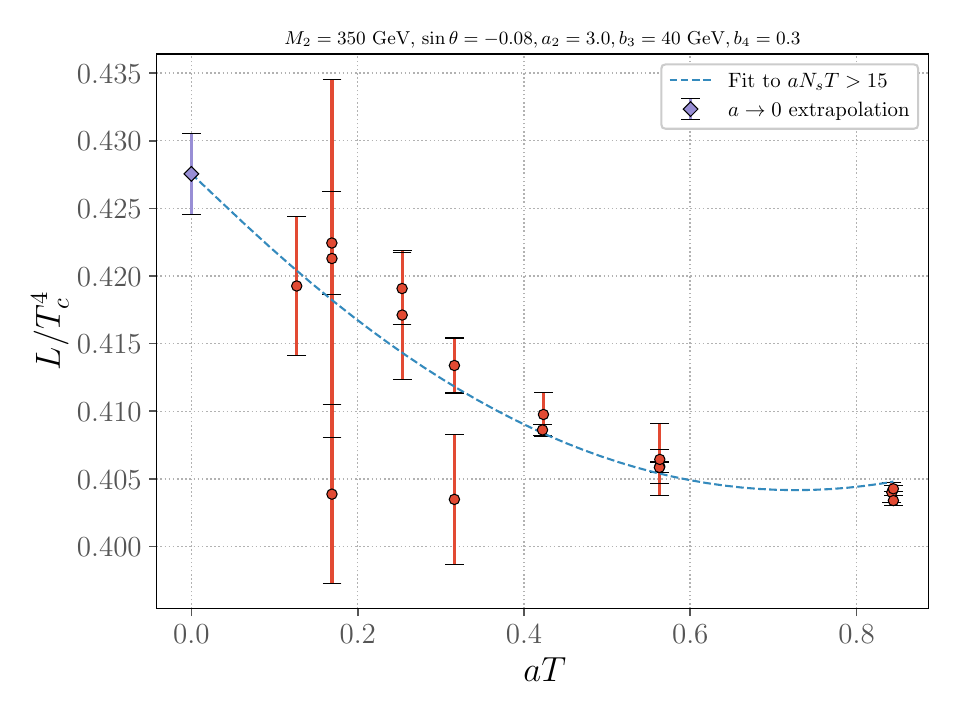}}
		\subfloat[Method 2: Using Eq.~(\ref{eq:latent_dSdT}).]{\includegraphics[width=0.5\textwidth]{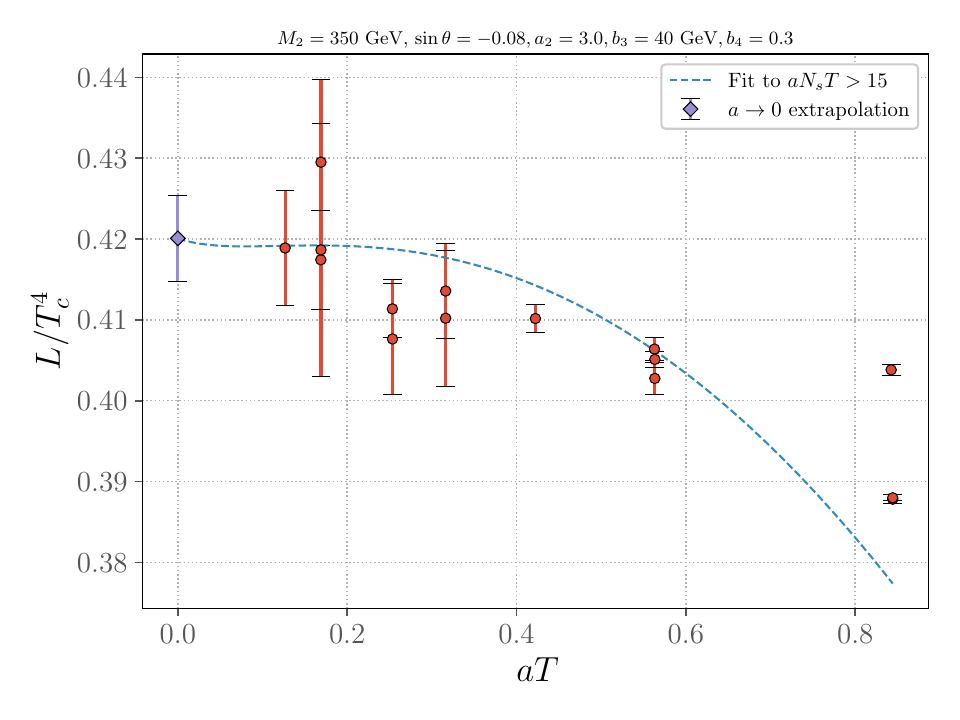}}
		\caption{Continuum extrapolations for the latent heat, computed using the two methods described in section~\ref{sec:strength}. The fitting ansatze are of the form described in the caption of Fig.~\ref{fig:BM3_spacing}.}
		\label{fig:BM3_latent}
	\end{figure*}

	The lattice theory gives physically meaningful results after removal of UV and IR cutoffs. Thus, we must take the continuum and infinite volume limits. The traditional approach involves performing simulations at different lattice spacings and volumes and extrapolating $a\rightarrow 0, V\rightarrow \infty$. Doing so requires considerable numerical effort that is impractical to implement for every parameter point studied in this work. For our purposes it suffices to perform simulations at small enough $a$ (large $\beta$) so that discretization effects are sufficiently small, {\it e.g.} of same order of magnitude or smaller as the expected errors from our 4D $\rightarrow$ 3D mapping. Similar considerations apply to finite-volume dependence.

	For the phase transition in 3D $\SU2$ + Higgs theory discretization errors are well under control for $\beta \gtrsim 20$ \cite{Kajantie:1995kf,Gould:2022ran}. In BSM theories one can expect more significant $\mathcal{O}(a)$ corrections if there are large couplings in the scalar potential -- as is often the case in first-order EWPT scenarios -- or if the BSM excitations are heavy comparable to the magnetic scale $g^2T$; see for example simulations in the two-Higgs doublet model \cite{Kainulainen:2019kyp}. To investigate cutoff dependence in our singlet extension we have performed a benchmark study in the following parameter point:
	\begin{align}
	\label{eq:BM3}
	\{ M_2, \sin\theta &, a_2, b_3, b_4 \} \nn 
    &= \{ 350 \text{ GeV}, -0.08, 3.0, 40 \text{ GeV}, 0.3 \}.
	\end{align}
	With these parameters we find the transition to be first order, in agreement with two-loop perturbation theory \cite{Niemi:2021qvp}\footnote{This is the benchmark point ``BM3'' from \cite{Niemi:2021qvp}}.

	Fig.~\ref{fig:BM3_volume} shows our results for the critical temperature and order parameter discontinuity as functions of the lattice side length, $N_s$ labeling the number of lattice sites in one direction, and at various values of $\beta$. The errors shown are statistical only and obtained with the blocked jackknife method. For fixed $\beta$ and for lattices larger than $aN_sT \gtrsim 15$ the results display a high degree of volume independence. The physical explanation is that $\SU2$ at high temperature has a mass gap of order $m \sim g^2T$, 
    and correlations over distances larger than $1/m$ are exponentially small. The hypercharge field remains nonperturbatively massless and has long-range correlations \cite{Kajantie:1996qd}, but our results do not indicate sensitivity to these effects.

	Turning next to lattice spacing dependence, we have data from a broad range of $\beta$ values ranging from $12$ to $80$. Using $\bar{g}^2 \approx 0.4$ this range translates to $aT \in (0.13, 0.83)$. The continuum limit can be inferred by extrapolating the large-volume results to $1/\beta \rightarrow 0$. For small enough $a$ the dominant lattice spacing dependence originates from missing corrections to the lattice-continuum relations (see appendix~\ref{sec:lattice-counterterms}). Departure from the continuum limit has a power series expansion in $a$, with possible logarithmic corrections at each order. 
	
	For the benchmark point (\ref{eq:BM3}) we have enough data for quadratic fits, shown in Fig.~\ref{fig:BM3_spacing} for $T_c$ and the condensate discontinuity. Fig.~\ref{fig:BM3_latent} shows continuum extrapolations for the latent heat $L$ as computed with the two methods introduced above.\footnote{Volume dependence of $L/T_c^4$ is similar to that of $v/T_c$; here we only show points with $aN_s T > 15$.} 
	The latent heat has considerably larger statistical errors than $T_c$ or $v/T_c$. At least for method two, this situation is to be expected since the measurement involves more condensates than just $\langle \he\phi\phi\rangle$. Despite the relatively large uncertainty in $L/T_c^4$, the $a\rightarrow 0$ extrapolated values from both methods agree well within error bars.

    \begin{table}[!htb]
		\begin{center}
			\begin{tabular}{|c|c|c|c|c|}
				\hline
				& $\Tc / \mathrm{GeV}$ & $v / \Tc$ & $L/\Tc^4$ (1) & $L/\Tc^4$ (2) \\
				\hline
				$\beta = 40, N_s = 80$ & 126.997(14) & 1.1282(7) & 0.4191(27) & 0.4291(3) \\
				$\beta \rightarrow \infty, N_s \rightarrow \infty$ & 127.053(9) & 1.1520(7) & 0.4285(30) & 0.4201(53) \\
				\hline
			\end{tabular}
		\end{center}
		\caption{Results at the benchmark point (\ref{eq:BM3}). The parentheses denote statistical error in the last digits. The two values of latent heat correspond to Eqs.~(\ref{eq:latent_prob}) and (\ref{eq:latent_dSdT}) respectively.
        Numbers on the second row are extrapolations to the continuum and infinite volume as described in the text.
        }
		\label{table:BM3}
	\end{table}
	We also show in table~\ref{table:BM3} the results from our simulation at fixed UV and IR cutoffs, at $\beta=40, N_s=80$. The numbers are well within $5\%$ of the $a\rightarrow 0$ extrapolated values, with the largest deviation being in the latent heat. This level of precision is more than reasonable for phenomenological purposes and should also be sufficient for benchmarking analytical calculations, at least until perturbative corrections beyond two-loop become available. For remainder of this paper we shall work at fixed $\beta=40, N_s=80$.

	\section{First-order transitions in the singlet model}
    \label{sec:results_heavy}

	\begin{figure*}[!htb]
		\subfloat[$a_2 = 2.5$]{\includegraphics[width=0.333\textwidth]{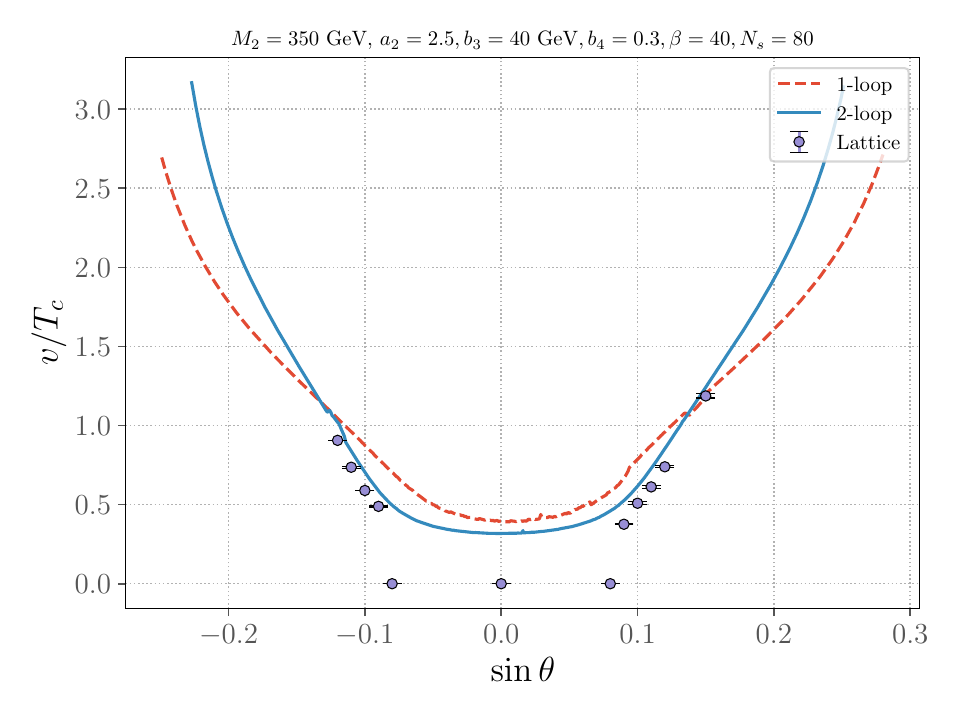}}
		\subfloat[$a_2 = 3.0$]{\includegraphics[width=0.333\textwidth]{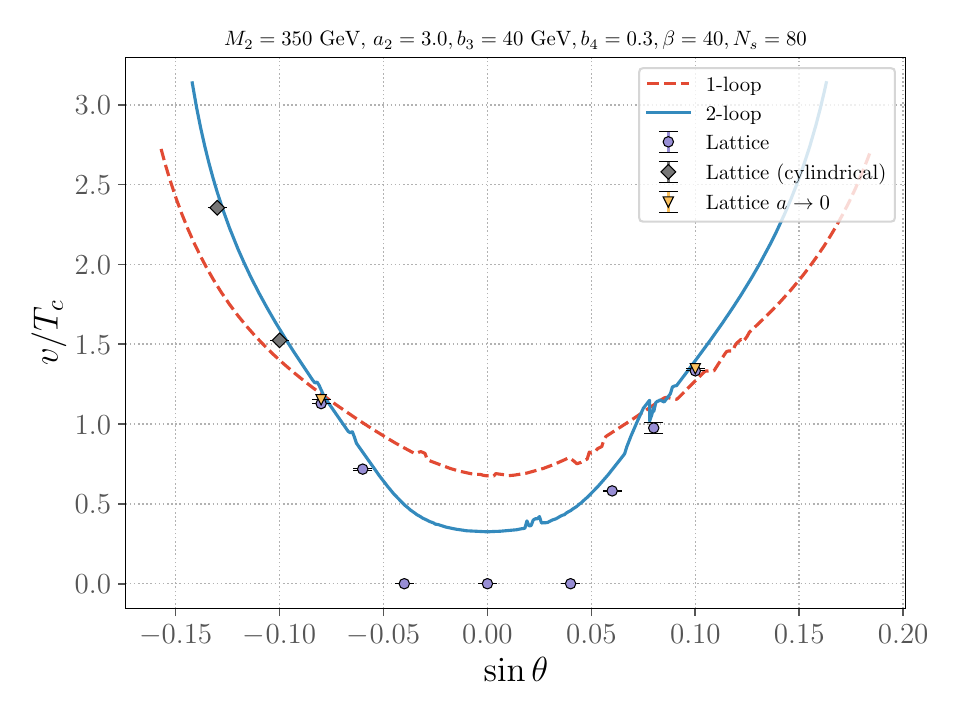}}
		\subfloat[$a_2 = 3.5$]{\includegraphics[width=0.333\textwidth]{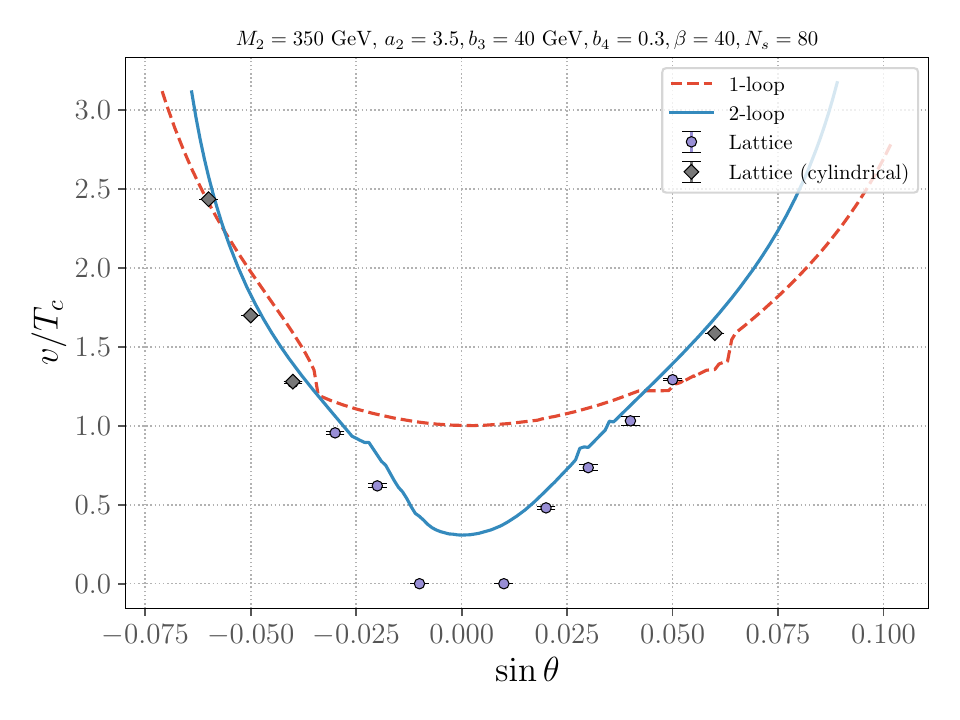}}
		\caption{Order parameter discontinuity, Eq.~(\ref{eq:higgs-vev}), as function of $\sin\theta$. For each $a_2$ the transitions at small $|\sin\theta|$ are of crossover type as indicated by $v/T_c = 0$. The simulations at at $\beta = 40, N_s = 80^3$; for $a_2 = 3.0$ we also show the $\beta, N_s\rightarrow \infty$ extrapolations at two values of $\sin\theta$. For very weak transitions we have utilized volumes up to $128^3$ to distinguish crossovers from weak first-order transitions.
		}
		\label{fig:BM3_st_vByT}
	\end{figure*}

	\begin{figure*}[!htb]
		\subfloat[$a_2 = 2.5$]{\includegraphics[width=0.333\textwidth]{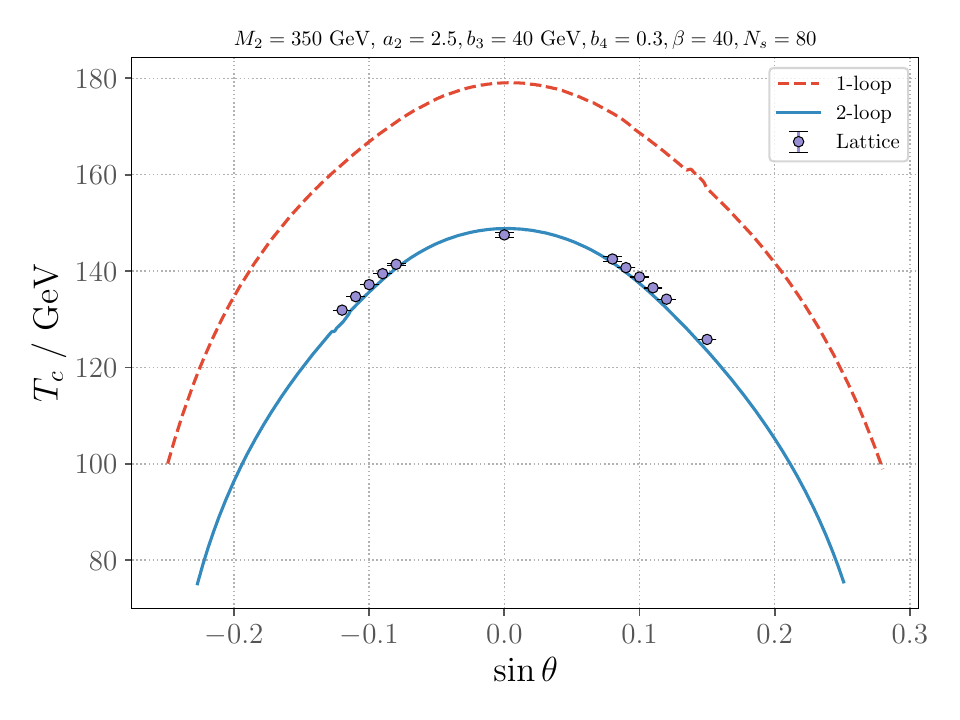}}
		\subfloat[$a_2 = 3.0$]{\includegraphics[width=0.333\textwidth]{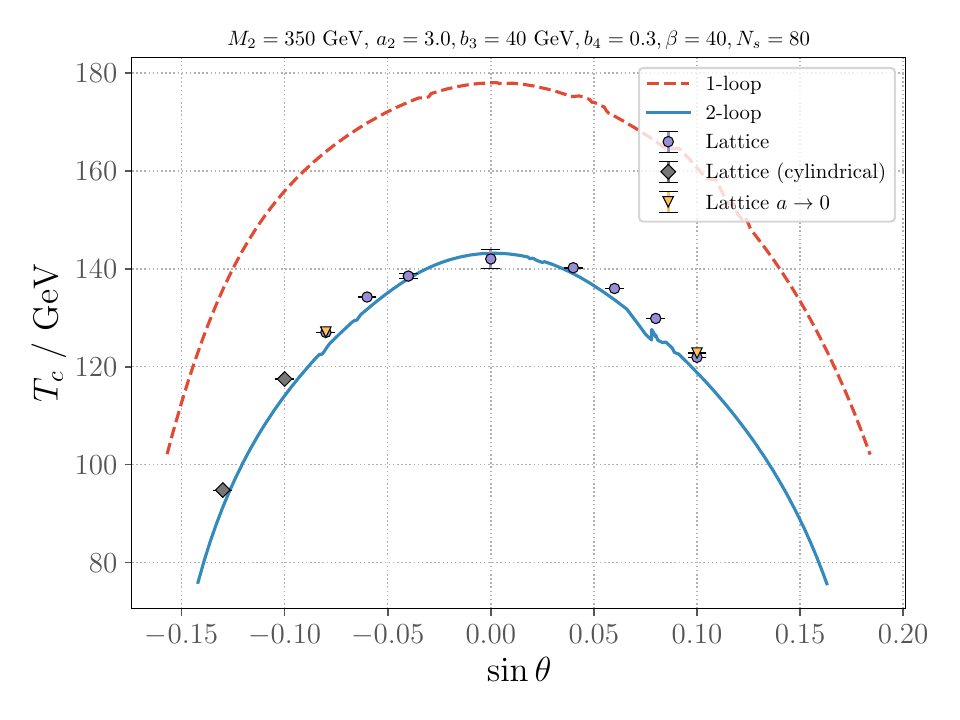}}
		\subfloat[$a_2 = 3.5$]{\includegraphics[width=0.333\textwidth]{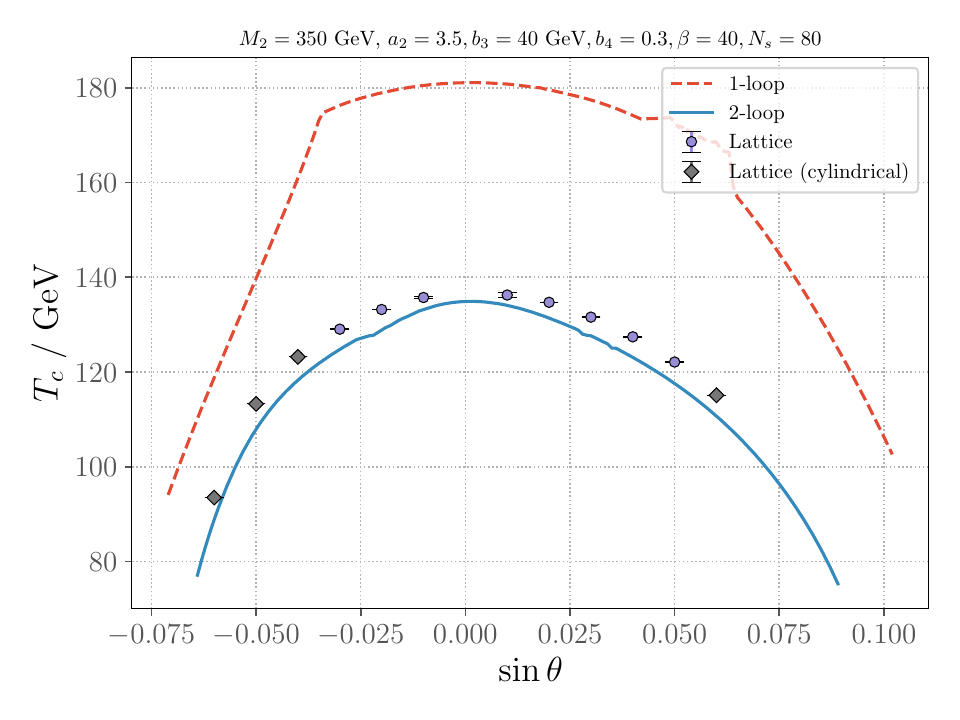}}
		\caption{As in Fig.~(\ref{fig:BM3_st_vByT}) but for the critical temperature.}
		\label{fig:BM3_st_Tc}
	\end{figure*}
	
	\begin{figure*}[!htb]
		\subfloat[$a_2 = 2.5$]{\includegraphics[width=0.333\textwidth]{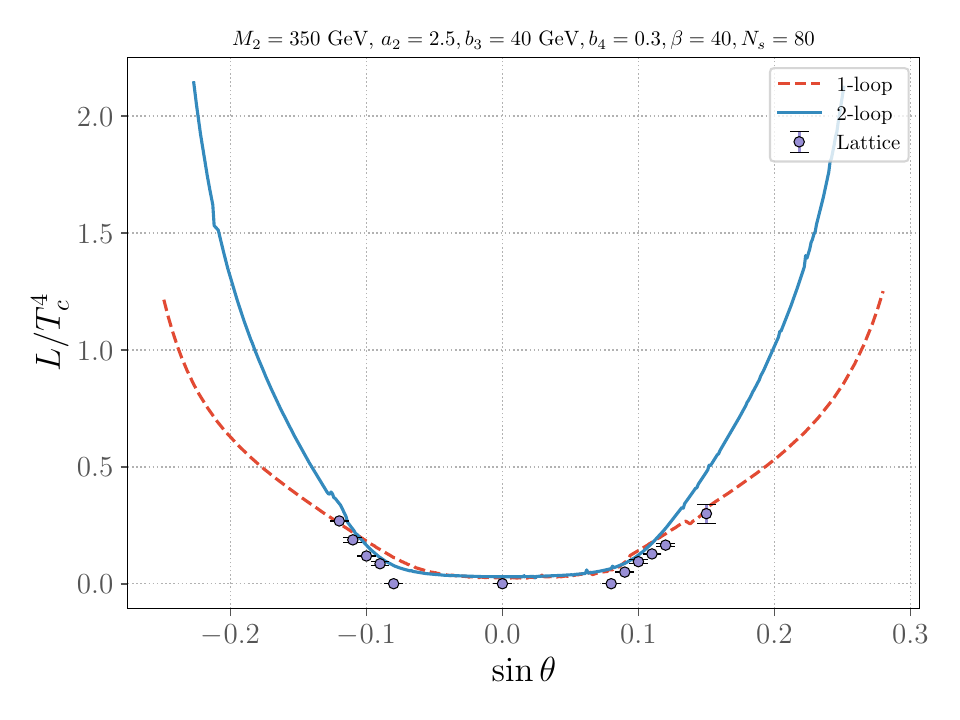}}
		\subfloat[$a_2 = 3.0$]{\includegraphics[width=0.333\textwidth]{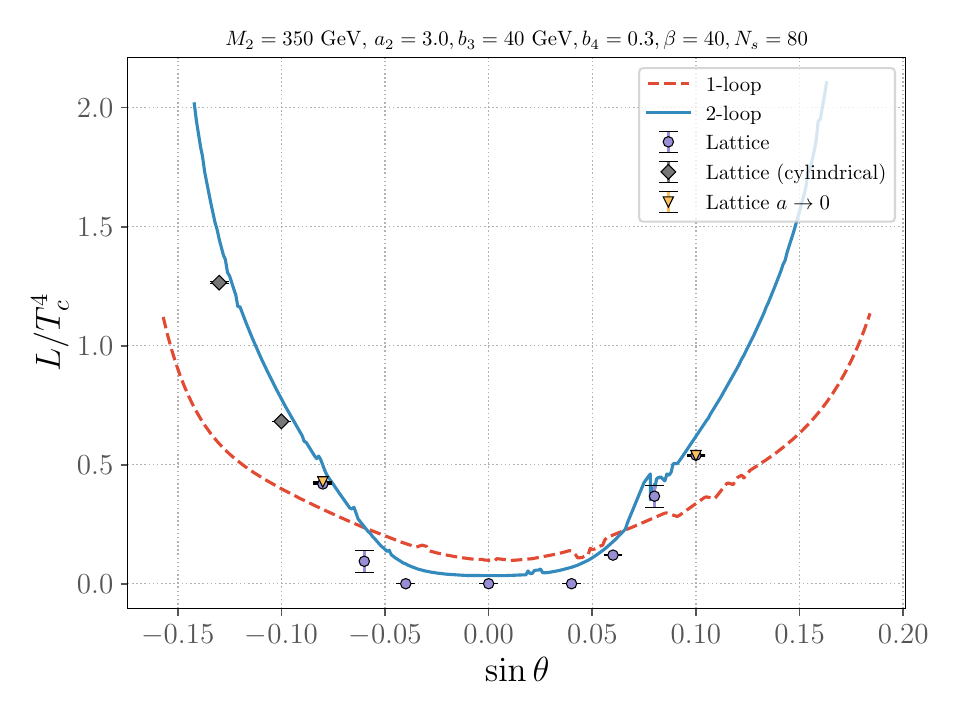}}
		\subfloat[$a_2 = 3.5$]{\includegraphics[width=0.333\textwidth]{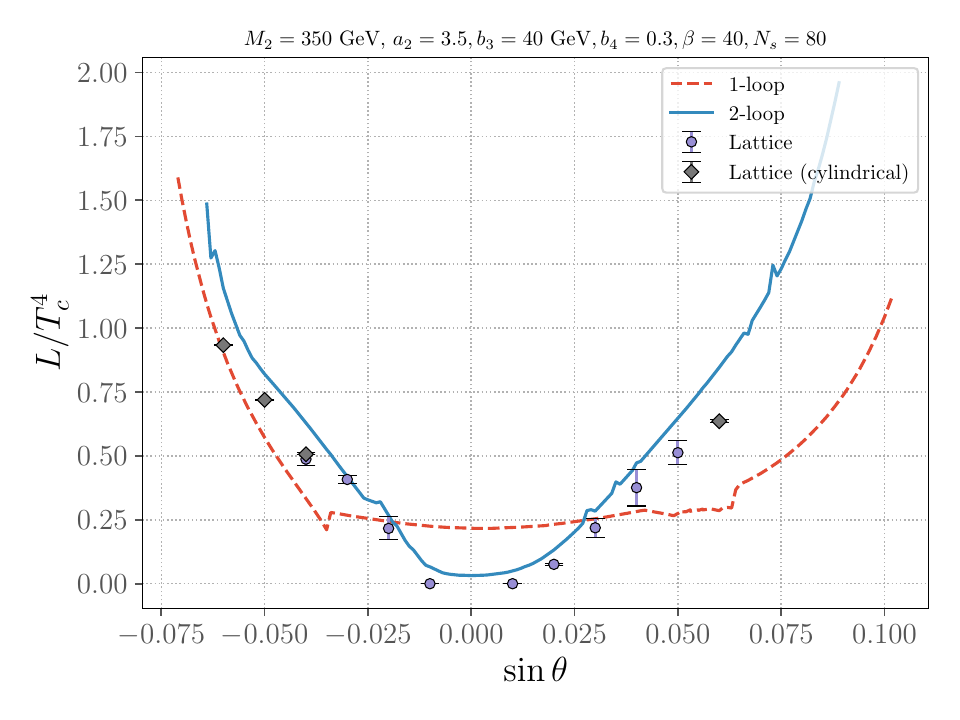}}
		\caption{As in Fig.~(\ref{fig:BM3_st_vByT}) but for the latent heat. The majority of points here use the method of Eq.~(\ref{eq:latent_prob}). Points obtained from ``cylindrical'' lattices instead use Eq.~(\ref{eq:latent_dSdT}).}
		\label{fig:BM3_st_latent}
	\end{figure*}
	
	What are the conditions for having a first-order EWPT in the singlet-extended SM? The minimal SM itself \textit{would} have a first-order transition if the Higgs self-coupling $\lambda$ was smaller by roughly a factor of two \cite{Kajantie:1996mn,Kajantie:1995dw,Csikor:1998eu}.
    In the singlet-extended model there are a few main mechanisms for turning the SM crossover into a proper phase transition: (i) Effective decoupling of the BSM degree of freedom but with sufficiently large (negative) renormalization of the Higgs quartic coupling, and (ii) having a more complicated vacuum structure so that a free-energy barrier can exist already in the tree-level scalar potential (see discussion in section~\ref{sec:model}). There is also option (iii) in which the other (non-Higgs) direction in the effective potential play little to no role, yet the new scalar is either too light or too strongly coupled to the Higgs that describing the transition in terms of decoupling is not reliable.

	\begin{figure*}[!htb]
		\subfloat[Critical temperature.]{\includegraphics[width=0.333\textwidth]{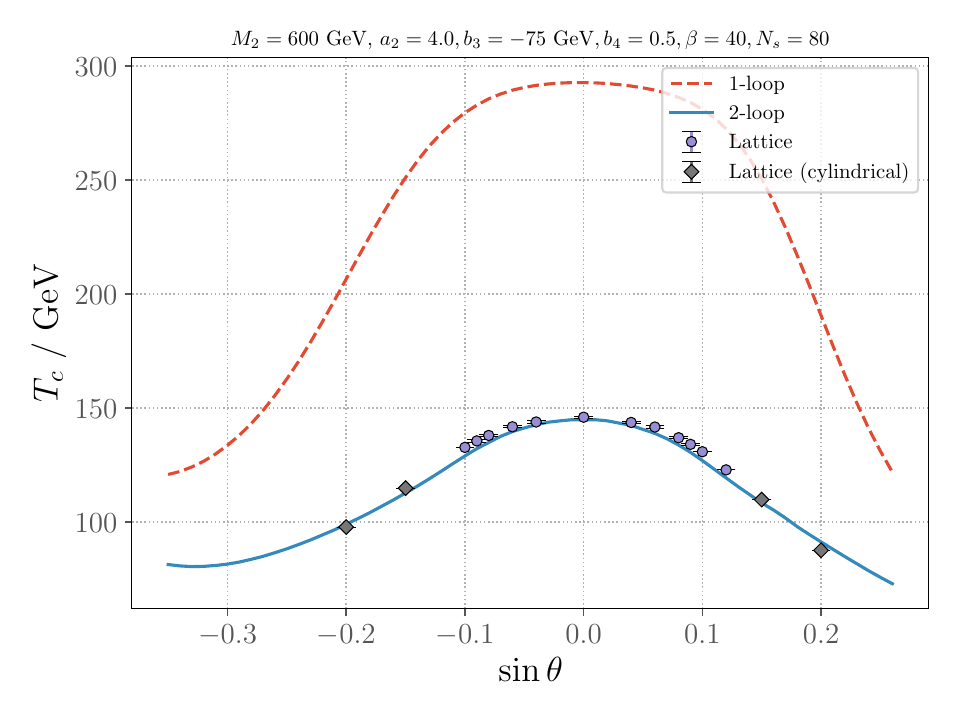}}
		\subfloat[Order parameter discontinuity.]{\includegraphics[width=0.333\textwidth]{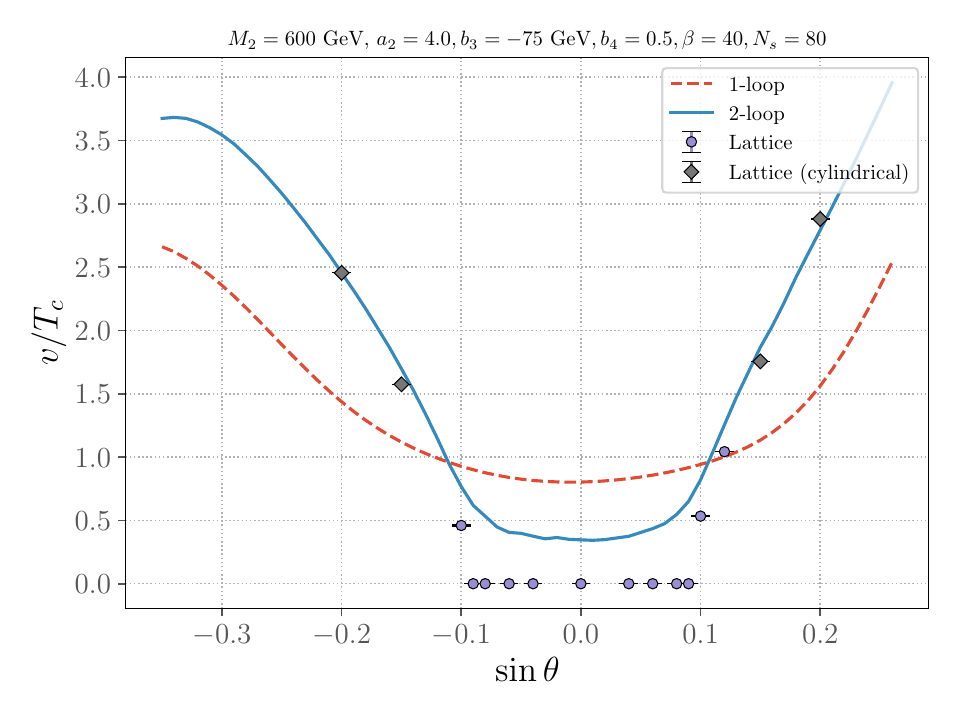}}
		\subfloat[Latent heat.]{\includegraphics[width=0.333\textwidth]{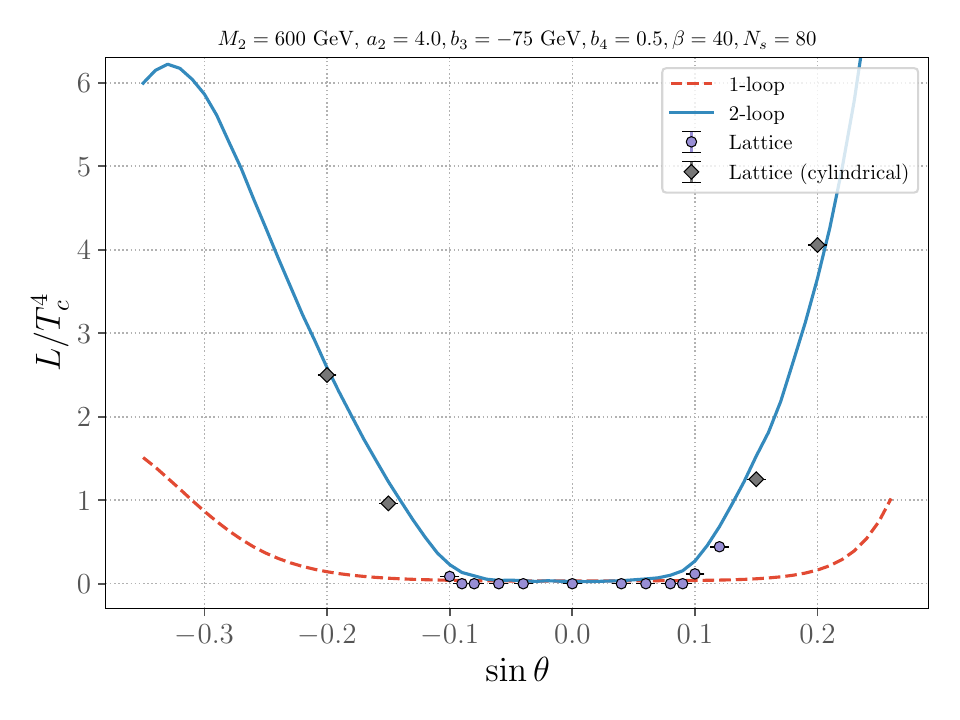}}
		\caption{Effect of $\sin\theta$ on the EWPT when the new particle is considerably heavier than the EW scale, $M_2 = 600$ GeV. Perturbative results (solid and dashed curves) are shown for comparison.}
		\label{fig:M600_vary_sintheta}
	\end{figure*}
	
	These scenarios have been analyzed perturbatively in numerous previous studies, and we refer the reader to Ref.~\cite{Ramsey-Musolf:2019lsf} for a summary and references. For our present purposes, we note that option (i) was studied in Ref.~\cite{Gould:2019qek} by perturbatively integrating out the singlet field and inferring the phase diagram from existing lattice results for the resulting $\SU2$ + Higgs effective theory (the $\grU1$ field was not considered). In this way regions of first-order EWPT could be mapped out. However, the reduction to $\SU2$ + Higgs was found to be unreliable particularly for stronger phase transitions. In the latter regime, associated with larger values of the effective triscalar couplings, the Wilson coefficient of the generated $(\phi^\dag \phi)^3$ operator becomes significantly larger than in the pure SM case, raising concerns about the reliability of the $\SU2$ + Higgs EFT. Including the singlet field explicitly in the simulations, as we do here, overcomes this limitation.
	
	In the generic singlet model without $Z_2$ symmetry, effect (ii) arises naturally through cubic interactions in the tree-level potential. In the context of perturbation theory, one may na\"ively expect that a tree-level barrier would guarantee a first-order EWPT. We now show that this expectation is not realized in the nonperturbative context.
    Letting $\theta$ denote the doublet-singlet mixing angle, which becomes non-zero in the presence of the cubic Higgs portal interaction $S\phi^\dag\phi$, we find that for small but nonzero $|\sin\theta|$ the transition can remain crossover. To study this situation we have performed simulations at different $\sin\theta$ with other parameters fixed according to Eq.~(\ref{eq:BM3}). We also repeated this analysis at different $a_2 = 2.5, 3.0, 3.5$.

	Our results for the Higgs condensate discontinuity are collected in Fig.~\ref{fig:BM3_st_vByT} together with perturbative estimates at 1- and two-loop order.\footnote{Our conventions for perturbative calculations are as in ref.~\cite{Niemi:2021qvp}
    In short, one-loop refers to $\mathcal{O}(g^3)$ calculation and has thermal masses only at one-loop, whereas two-loop is a full $\mathcal{O}(g^4)$ calculation. For both computations we use same one-loop corrected matching between input parameters and the \MSbar renormalized parameters in the zero-temperature theory.} 
	For small $|\sin\theta|$ we do not observe discontinuous behavior in the order parameter (histograms do not develop two-peak structure as in Fig.~\ref{fig:hgrams}), indicating absence of a first-order transition. These points are indicated by $v/T_c = 0$ in Fig.~\ref{fig:BM3_st_vByT}. The EWPT strength is also seen to be quite sensitive to the value of $a_2$, with smaller $|\sin\theta|$ being sufficient for first order transition at larger $a_2$. As a concrete example, the na\"ive baryon number preservation criterion \cite{Morrissey:2012db} $v/T \gtrsim 1$ for EW baryogenesis is satisfied for $|\sin\theta| \gtrsim 0.08$ at $a_2 = 3.0$, while at $a_2 = 3.5$ already $|\sin\theta| \gtrsim 0.03$ is sufficient. Results for $T_c$ and the latent heat are shown in Figures \ref{fig:BM3_st_Tc} and \ref{fig:BM3_st_latent}.
	
	The crossover behavior at small $|\sin\theta|$ is in stark contrast to perturbative results that always predict a first order transition, although two-loop corrections do bring the discontinuity closer to zero in the crossover region. In the context of the dimensionally reduced effective field theory, two-loop perturbation theory is qualitatively reliable once the transition becomes first order. Indeed, figures \ref{fig:BM3_st_vByT}-\ref{fig:BM3_st_latent} show that the two-loop EFT predictions agree fairly closely with our nonperturbative results. Discrepancies are, unsurprisingly, largest in the $a_2 = 3.5$ case where the difference from lattice results reaches $30\%$ for $L/T_c^4$. It is also clear that one-loop estimates for all quantities differ significantly from the nonperturbative values.

    We have also performed simulations at larger $M_2 = 600$ GeV. One expects that as the BSM excitation becomes heavier, a compensating increase in $|\sin\theta|$ or $a_2$ is needed to prevent decoupling and remain in the first order EWPT-viable parameter space. Previous studies using conventional one-loop perturbative thermal computations support this expectation
    \cite{Curtin:2014jma,Chala:2016ykx,Kurup:2017dzf}. Our lattice results at $M_2 = 600$ GeV and $a_2 = 4$ are shown in Fig.~\ref{fig:M600_vary_sintheta} where we again vary the mixing angle. The behavior is similar to the $M_2 = 350$ GeV scenarios above, and the two-loop predictions  once again agree closely with the lattice results. However, we observe a  a larger discrepancy between 1- and two-loop calculations. A relatively large $|\sin\theta| \gtrsim 0.1$ is needed for a first order transition, beyond which the EWPT strength increases rapidly with the mixing angle.

    We point out that the $M_2 = 600$ GeV case may be pushing our high-$T$ EFT approach to its limits. The formal condition $M_2 \leq \pi T$ is not satisfied. Thus, higher-order corrections to the EFT, and consequently to the lattice action, can become important (see discussions in sections \ref{sec:EFT} and \ref{sec:lattice-theory}). A method for estimating accuracy of the EFT, based on dimension five and six operators, was outlined in ref.~\cite{Niemi:2021qvp}. Using this approach, we have verified that the error from neglecting these operators in the EFT in our $M_2 = 600$ GeV points is less than $5\%$ for $|\sin\theta| < 0.1$. At $|\sin\theta| = 0.2$ the error reaches $10\%$, meaning that in these points, the 3D lattice approach is an order of magnitude less accurate than in the minimal SM \cite{Kajantie:1995dw}.

	Note that we have performed our perturbative calculations with the high-$T$ effective potential constructed in \cite{Niemi:2021qvp} that includes all two-loop diagrams within the 3D EFT (\ref{sec:EFT}). The perturbative potential contains residual dependence on the \MSbar renormalization scale, formally a three-loop effect. All our perturbative results have been obtained at fixed scale $\bar\mu = T$.
    It has recently been argued that a reorganization
	of the loop expansion may be necessary for consistently describing first order phase transitions \cite{Ekstedt:2022zro,Gould:2023ovu}.\footnote{This reorganization is in addition to standard thermal mass resummations that are handled in the 4D $\rightarrow$ 3D step. Our perturbative approach corresponds to the direct minimization method discussed in appendix B of \cite{Gould:2023ovu}.}
	Although our two-loop results are in good agreement with the nonperturbative data, it would certainly be interesting to repeat the perturbative analysis using these new methods.

	Finally, we emphasize that that while in all cases studied here the $|\sin\theta|\rightarrow 0$ limit shows an absence of first-order EWPT, this behavior does not rule out first-order transitions in the entire $Z_2$ symmetric model. Indeed, our $|\sin\theta| = 0$ case does not even correspond directly with the $Z_2$ limit because we retain the cubic $ S^3$ interaction.  We also have not varied the $S^4$ coupling $b_4$ that affects two-step EWPT in particular \cite{Kurup:2017dzf} (see \cite{Niemi:2020hto} for analogous discussion in a triplet-Higgs extension).

	\section{Simulations with light singlet}
	\label{sec:small-mass}
	
	\begin{figure*}[!htb]
		\subfloat[Critical temperature.]{\includegraphics[width=0.333\textwidth]{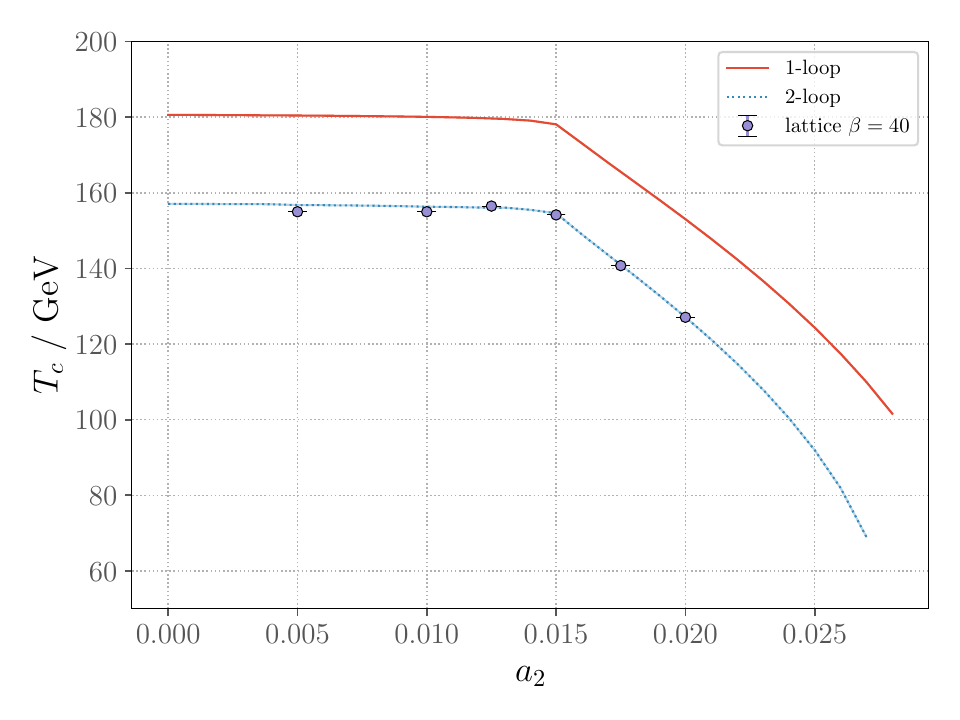}}
		\subfloat[Higgs condensate jump.]{\includegraphics[width=0.333\textwidth]{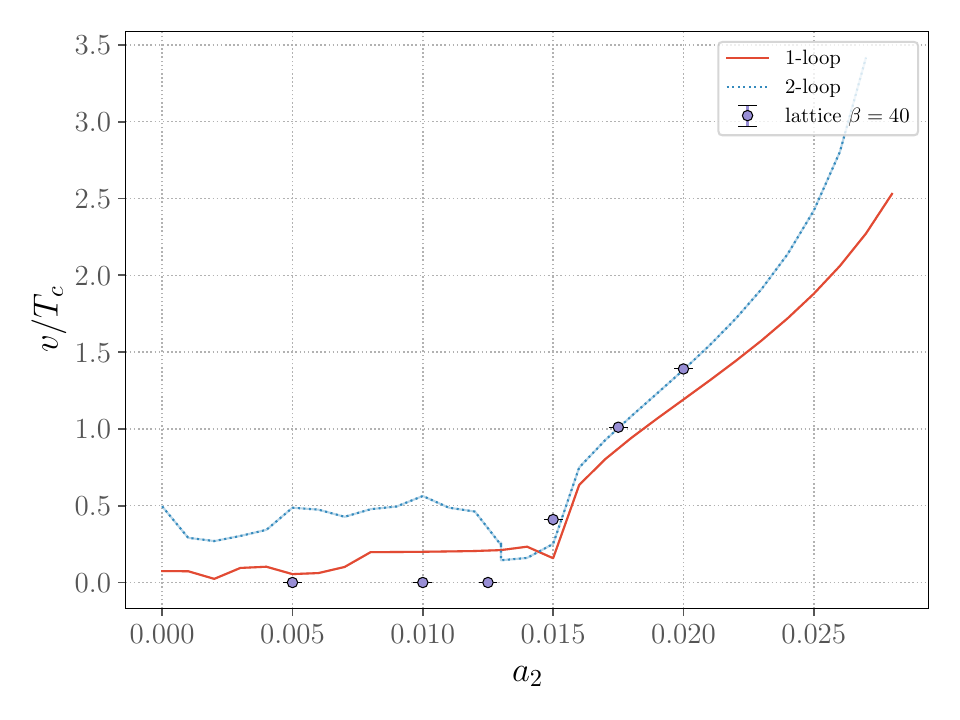}}
		\subfloat[Latent heat.]{\includegraphics[width=0.333\textwidth]{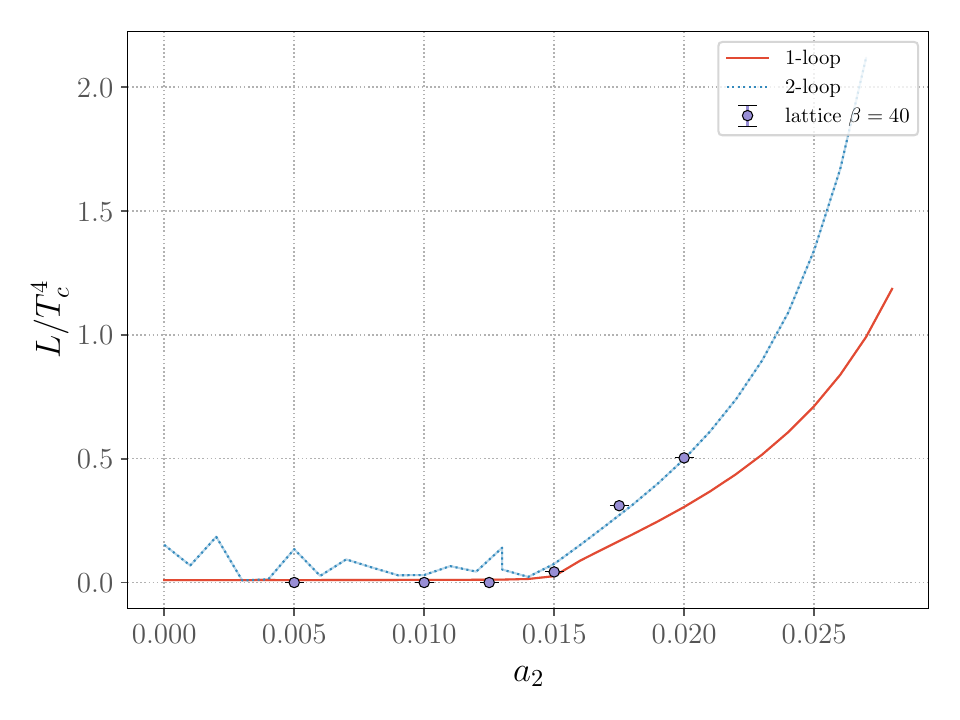}}
		\caption{Simulation results for $T_c$ and phase transition strength at $M_2 = 25$ GeV and different $a_2$. The other parameters were $\sin\theta = 0.01, b_4 = 0.0015, b_3 = 1$ GeV. We have not studied the region $a_2 \gtrsim 0.03$ as there the $T=0$ EW vacuum is not stable with respect to another, deeper minimum where the Higgs mechanism is absent.}
		\label{fig:small_mass_a2}
	\end{figure*}

	It is interesting to consider whether the singlet-extended model can accommodate a first-order EWPT given a light BSM excitation, $M_2 \leq M_1 / 2$ ($M_1 = 125$ GeV is the SM Higgs mass). The presence of such light scalar would open a new window for collider phenomenology through ``exotic'' decays of the Higgs into two new bosons, schematically $h_1 \rightarrow h_2 h_2$. Phase transitions in this small-mass region were studied in \cite{Kozaczuk:2019pet} at one-loop level where it was argued that the EWPT strength is predominantly controlled by the $S^2 \he\phi\phi$ interaction (see also \cite{Carena:2019une, Carena:2022yvx}).

 In the small-$\theta$ limit\footnote{The mixing angle in \cite{Kozaczuk:2019pet} is related to ours by $\cos\theta \rightarrow -\sin\theta$, and their $h_1$ scalar is our $h_2$.} , $a_2$ directly fixes also the interaction strength relevant for $h_1 \rightarrow h_2 h_2$ decays \cite{Kozaczuk:2019pet}. The argument set forth in \cite{Kozaczuk:2019pet} is that the requirement of a  successful strongly first-order EWPT sets a lower bound on $a_2$, while experimental searches for exotic Higgs decays can set an upper bound.  
 
    According to one-loop scans \cite{Kozaczuk:2019pet, Carena:2019une}, the small-mass region can accommodate a strong first-order EWPT even at exponentially small $a_2$ coupling, ranging from $10^{-3}$ to $10^0$ depending on the mass. This situation differs starkly from the heavy-singlet region $M_2 > 125$ GeV where strong transitions are predominantly associated with $a_2 > 1$ \cite{Curtin:2014jma,Chala:2016ykx,Kurup:2017dzf,Gould:2019qek}.
	
	To assess the light $M_2$ regime
	we have simulated the EWPT at $M_2 = 25$ GeV and studied the effect of $a_2$; the results for one such benchmark scenario are shown in Fig.~\ref{fig:small_mass_a2}. The transition is a crossover for $a_2 \lesssim 0.015$ and first order for larger $a_2$. Perturbation theory again becomes reliable once the transition becomes first order, and the agreement with lattice data is even better than in the benchmark cases presented earlier. One-loop results become increasingly unreliable as $a_2$ increases, even though $a_2$ here is small compared to other relevant parameters such as gauge and top Yukawa couplings. 
	
	The two-loop analysis is notably unreliable in the region where lattice simulations reveal a crossover transition. This situation -- visible as  ``jagginess'' in the two-loop curves for EWPT strength -- is likely caused by IR divergent scalar loops (which are absent at one-loop order) invalidating the minimization of the effective potential near $T_c$; see \cite{Laine:1994zq, Kajantie:1995kf, Niemi:2020hto, Gould:2023ovu} for related discussions. However, we do not have a clear understanding of why the issue is more predominant here than in our other parameter-space points.

	The simulations at small $M_2$ turned out to be technically challenging. For instance, in the $a_2 = 0.02$ case on a $\beta = 40, N_s = 80$ lattice, the probability boost from the multicanonical algorithm had to be of order $e^{190}$ for the simulation to tunnel, and obtaining a good weight function for this took over $10000$ CPU hours. The mixed-phase method (described at the end of section~\ref{sec:find-Tc}) was numerically more manageable, and we used it for collecting the final data. We also attempted simulations at $a_2 = 0.025$ which presumably yields a much stronger transition. Unfortunately both methods failed to give a reliable estimate for the transition temperature in this point. Choosing a more optimized order parameter - {\it e.g.} a linear combination of $\phi$ and $S$ condensates - for the multicanonical algorithm could make the simulations easier in this region, but we have not attempted this. See \cite{Laine:1998qk} for related considerations. We defer a more comprehensive treatment of the light $M_2$ region to future work.

    \section{Model-phenomenological considerations}
    \label{sec:pheno}

    \begin{figure}[!htb]
         \includegraphics[width=\columnwidth]{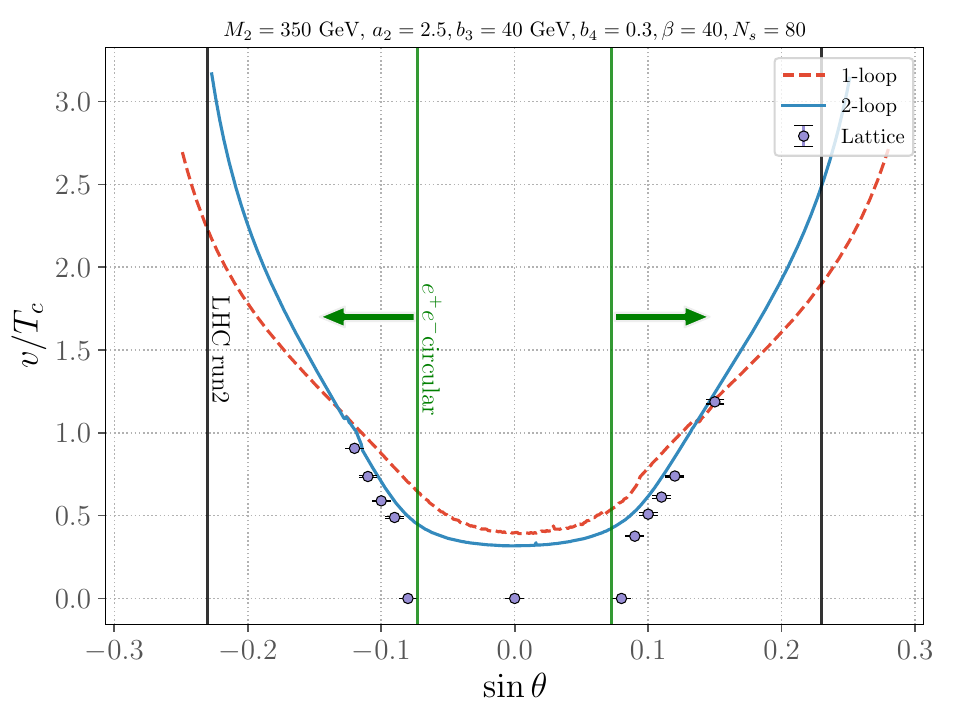}
         \caption{Higgs condensate discontinuity as a function of $\sin\theta$, with  $M_2 = 350\ \text{GeV}, a_2=2.5, b_3=40\ \text{GeV}, b_4 = 0.3$. 
         Red dashed and blue curves give
         one-loop and two-loop perturbative results, respectively.  Lattice results are obtained at $\beta = 40$ and a volume of $80^3$.
         Points at small $|\sin\theta|$, where the order parameter discontinuity vanishes, indicate a crossover transition.  Sensitivities of current and future Higgs measurements  are indicated by vertical lines. The current limit from LHC-Run2 is shown by the black solid lines. The projected constraint from $\sqrt{s}=240$ GeV, $5.6 \ \mathrm{ab}^{-1}$ CEPC or $\sqrt{s} = 240$ GeV, $5 \ \mathrm{ab}^{-1}$ FCC-ee is given by the green solid lines.}
         \label{fig:sintheta-pheno}
    \end{figure}

    Thus far, we have utilized lattice simulations to (a) identify the parameter space boundary between a crossover and first-order EWPT and (b) assess the quantitative reliability of perturbative computations in the first-order EWPT region. We now consider the phenomenological implications when the singlet-like scalar $h_2$ is either heavier or  lighter than the SM-like scalar $h_1$. For the heavy $h_2$ regime,
    figures \ref{fig:BM3_st_vByT} and \ref{fig:M600_vary_sintheta} show that, in the parameter space studied here, first-order transitions are associated with nonzero value of the mixing angle $\theta$ and that the transition grows stronger when $|\sin\theta|$ increases. At the same time, collider measurements for Higgs boson signal strengths put upper bounds on $|\sin\theta|$ (see for instance \cite{Profumo:2014opa}).

    It is interesting to assess the present and future collider sensitivities to $\sin\theta$ in light of our lattice results. To that end, we illustrate these sensitivities in Fig.~\ref{fig:sintheta-pheno} for the $M_2 = 350$ GeV, $a_2 = 2.5$ case from Fig.~\ref{fig:BM3_st_vByT}.The vertical black and green lines indicate, respectively, the present LHC bounds and prospective future circular $e^+e^-$ collider reach.
    We obtain the former from a combined fit to
    The current experimental restrictions on the mixing angle from Higgs boson signal strengths ($\mu$) are derived from the combined fit to the latest ATLAS \cite{ATLAS:2022vkf} and CMS \cite{CMS-PAS-HIG-19-005} results, which are $\mu=1.05 \pm 0.06$ and $\mu=1.02^{+0.07}_{-0.06}$ respectively. Employing the $\chi^2$ method described in \cite{Profumo:2014opa}, the $95\%$ C.L. limit on mixing angle is $|\sin\theta|\lesssim 0.23$.
    We also collect the associated uncertainties in projected Higgs signal rate of various future $e^+ e^-$ colliders from \cite{deBlas:2019rxi}. We use the same $\chi^2$ fit method to determine the corresponding projected limits on $\sin\theta$, which are listed in table~\ref{table:collider_sensitivity}. As the sensitivities of CEPC and FCC-ee are similar, we use \lq\lq $e^+ e^-$ circular\rq\rq\,  as a reference for both collider options. The arrows pointing to the left and right indicate
    the corresponding prospective $95\%$ C.L. $e^+ e^-$ circular collider sensitivity region.

    Note that these projections do not include constraints from electroweak precision observables (EWPO). As illustrated in \cite{Huang:2017jws}, EWPO yield constraints in the $\sin\theta$-$M_2$ plane, whereas the Higgs signal strength $\sin\theta$ sensitivities are $M_2$-independent for a heavy singlet-like scalar. Thus, the location of the vertical lines in Fig.~\ref{fig:sintheta-pheno} will be the same for any of the other heavy singlet-like scalar benchmark cases discussed above. Importantly, we observe that prospective future circular colliders will achieve sensitivity most of the first-order EWPT-viable $\sin\theta$ parameter space\footnote{The projected linear $e^+e^-$  collider sensitivities are somewhat weaker, as indicated in Table~\ref{table:collider_sensitivity}. }. Moreover, one- and two-loop perturbative results indicate the existence of such parameter space beyond the future $e^+e^-$ circular collider reach, whereas our lattice results indicate much, if not most, of this $\sin\theta$-inaccessible region is associated with crossover transitions. In short, our study indicates that the future circular colliders will be sensitive to a relatively greater portion of the first-order EWPT parameter space than implied by previous perturbative studies.

    \begin{table}[htbp]
    \centering
    \begin{tabular}{|l|c|c|c|c|}
    \hline
             & $\sqrt{s}$ & $\mathcal{L}_\text{int}$ & $\mathcal{P} [e^-/e^+]$ & Sensitivity\\
    \hline
    CEPC     & 240 GeV    & $5.6 \ \text{ab}^{-1}$   & $0\%/0\%$               & $|\sin\theta|\lesssim  0.0726$ \\
    \hline                                                                
    CLIC     & 380 GeV    & $1.0 \ \text{ab}^{-1}$   & $-80\%/0\%$             & $|\sin\theta|\lesssim  0.1155$ \\
             & 3.0 TeV    & $5.0 \ \text{ab}^{-1}$   & $-80\%/0\%$             & $|\sin\theta|\lesssim  0.0541$ \\
    \hline                                                            
    FCC-ee   & 240 GeV    & $5.0 \ \text{ab}^{-1}$   & $0\%/0\%$               & $|\sin\theta|\lesssim  0.0728$ \\
    \hline                                                                        
    ILC      & 250 GeV    & $2.0 \ \text{ab}^{-1}$   & $-80\%/+30\%$           & $|\sin\theta|\lesssim  0.1098$ \\
    \hline
    \end{tabular}
    \caption{Sensitivity to $\sin \theta$ from selected $e^+ e^-$ colliders. The respective center-of-mass energy $\sqrt{s}$, integrated luminosity $\mathcal{L}_{\text{int}}$, and polarization $\mathcal{P}$ are provided.}
    \label{table:collider_sensitivity}
    \end{table}

    We now consider the light $h_2$ regime. In section~\ref{sec:small-mass} we discussed simulations in parameter space where decay channel $h_1 \rightarrow h_2 h_2$ is kinematically allowed. The condition for this decay is roughly $M_2 \lesssim 60$ GeV. We currently have nonperturbative results only at $M_2 = 25$ GeV, which is not sufficient for making generic statements about the phase transition in this region. However, our lattice data strongly suggests that the two-loop effective potential is a valid tool for qualitatively describing the EWPT {when the transition is strongly first-order}. For $M_2 = 25$ GeV the two-loop predictions shown in Fig.~\ref{fig:small_mass_a2} display excellent quantitative as well as qualitative agreement with nonperturbative results. Motivated by this situation, we now take a somewhat distinct approach from rest of this paper and perform a broader perturbative study of the $M_2 \leq 60$ GeV region.

    \begin{figure}[!htb]
        \includegraphics[width=\columnwidth]{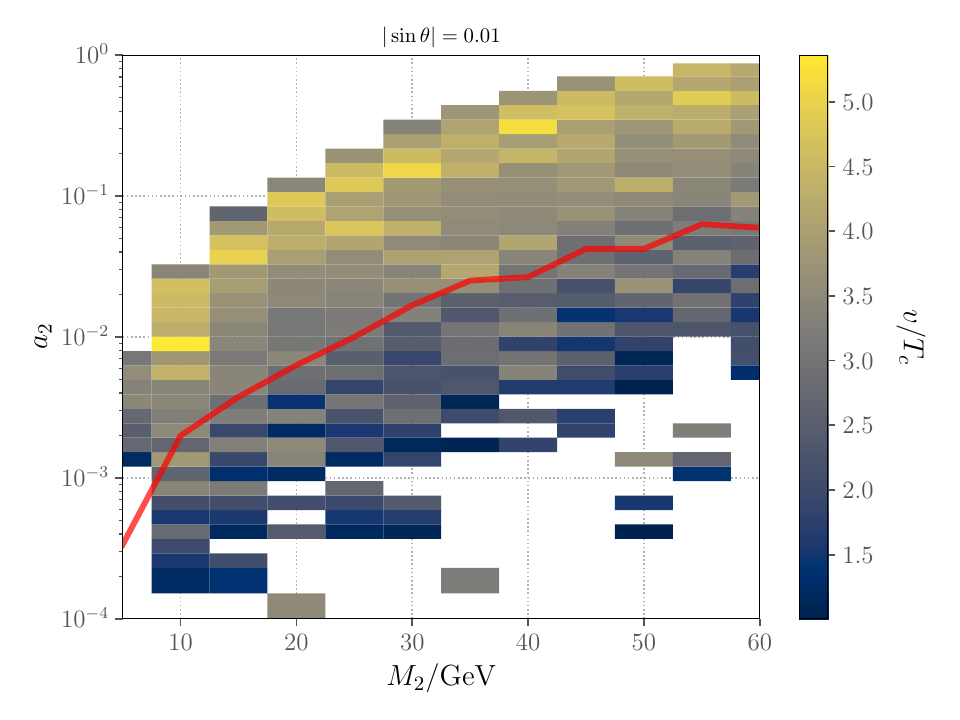}
        \caption{Result of a parameter scan over $M_2, a_2, b_3, b_4$, and fixed $|\sin\theta| = 0.01$, using two-loop perturbation theory. We show a projection to $(M_2, a_2)$ plane with each bin denoting the strongest transition found in the larger scan. The solid red curve is taken from the one-loop analysis of \cite{Kozaczuk:2019pet} and denotes a boundary of successful nucleation as described in the text. For points under the curve the cosmological phase transition never completes. As discussed in section~\ref{sec:small-mass}, lattice simulations at $M_2=25$ GeV indicate that for some parameter choices, the EW transition may be crossover even for values of $a_2$ that lie above the nucleation lower bound.
        }
        \label{fig:small_mass_scan}
    \end{figure}

    In this context, our main interest focuses on the strength of the EWPT in $(M_2, a_2)$ as these parameters have a decisive impact on exotic Higgs decay rate $\Gamma(h_1\to h_2 h_2)$. Following \cite{Kozaczuk:2019pet}, we consider small mixing angle $|\sin\theta = 0.01|$ and perform a logarithmic scan over the following parameters ranges:
    \begin{align}
        a_2 &\in [10^{-4}, 1], \quad \text{40 points} \nn
        b_3/v_0 &\in [10^{-4}, 1], \quad \text{50 points} \nn
        b_4 &\in [10^{-5}, 1], \quad \text{50 points},
    \end{align}
    with $v_0 = 246$ GeV.
    We vary the mass $M_2$  over the range $[5, 60]$ GeV with uniform spacing of $5$ GeV for total of $2.4$ million points. For $M_2 = 5$ GeV, we extend the lower bound for $b_4$ was extended to $10^{6.5}$. For each point, we increase the temperature in steps of $0.5$ GeV until the global minimum of the two-loop effective potential shifts to $0$ in the Higgs direction. Details of our potential are as in ref.~\cite{Niemi:2021qvp}. At the critical temperature we compute the condensate discontinuity $\Delta \langle \he\phi\phi \rangle$ and the ``physical'' Higgs VEV as defined in Eq.~(\ref{eq:higgs-vev}). We then project the results to $(M_2, a_2)$ plane, keeping only the largest $v/T_c$ value for each $M_2, a_2$ pair. The results appear in Fig.~\ref{fig:small_mass_scan}, where we only show transitions with $v/T_c > 1$. We have also excluded points where the EW minimum is not the global minimum of the $T=0$ potential (at one-loop).

The authors of Ref.~\cite{Kozaczuk:2019pet} performed an analogous scan using one-loop perturbation theory in the high-$T$ expansion. Our Fig.~\ref{fig:small_mass_scan} extends this analysis to two-loop order, with the limitation that our results do not include constraints from requiring successful nucleation. This distinction has important consequences for cosmology: if the nucleation rate is slow compared to the Hubble rate, the phase transition may never complete, leaving the universe stuck in the metastable high-temperature phase. The analysis in Ref.~\cite{Kozaczuk:2019pet} indicates that requiring that the transition completes implies a lower bound on $a_2$. The corresponding boundary is shown as a solid line in our Fig.~\ref{fig:small_mass_scan}, sampled with $M_2$ spacing of 5 GeV.\footnote{We thank Yanda Wu for providing the data for the boundary curve from his reproduction of results of Ref.~\cite{Kozaczuk:2019pet}.}
    The curve is parametrically less accurate than our parameter scan due to absence of two-loop contributions but we nevertheless expect it to be qualitatively correct, showing that points at too small $a_2$ do not correspond to a realistic cosmological phase transition.   

    The main message of Fig.~\ref{fig:small_mass_scan} is that  the nucleation condition typically puts a more stringent lower bound on the $a_2$ parameter than simply the requirement of a strong EWPT. This conclusion was also reached in \cite{Kozaczuk:2019pet}. 
    One may consider our results as a two-loop confirmation of this result, with one caveat:
    Being a two-dimensional projection of a five-dimensional parameter-space scan, Fig.~\ref{fig:small_mass_scan} effectively hides the effects of other parameters, especially $b_3, b_4$, on the EWPT. Indeed, as shown by our lattice results in Fig.~\ref{fig:small_mass_a2}, not all points for $a_2$ below the nucleation curve even lead to a first-order EWPT. In the perturbative computation, the successful nucleation condition, for $M_2=25$ GeV, $\sin\theta=0.01$, $b_4=0.0015$, $b_3=1$ GeV, is $a_2\gtrsim 0.0106$. Our lattice study indicates the transition is crossover for $a_2\lesssim 0.015$.

    As discussed in Ref.~\cite{Kozaczuk:2019pet}, for $M_2 \lesssim 60$ GeV the $a_2$ parameter cannot be arbitrarily large without violating experimental bounds on the Higgs decay width. A strong EWPT, in contrast, requires a sufficiently large $a_2$. This combination of considerations makes exotic Higgs decays a promising experimental probe of a singlet-catalyzed EWPT, particularly at small values of the mixing angle, a scenario that is otherwise hard to constrain experimentally.
    Our extended analysis here supports this key finding of Ref.~\cite{Kozaczuk:2019pet}.
    
	\section{Conclusions}
    \label{sec:conclude}
	
    In this paper, we have performed a detailed study of phase transition thermodynamics in the real-singlet extension of the Standard Model using nonperturbative lattice simulations. This nonperturbative method allows for a robust treatment of high-temperature IR physics relevant for the electroweak phase transition that can otherwise invalidate perturbative approaches. We hope that our overview of lattice concepts and techniques in sections~\ref{sec:lattice-theory} - \ref{sec:spacing} and the Appendices is useful for interested readers who are not already familiar with lattice field theory.
    
    The numerical cost of lattice simulations generally limits their utility for BSM models with many free parameters. Here we have 
    studied the phase transition as a function of the
    three parameters for collider phenomenology: the singlet-like scalar mass, $M_2$; doublet-singlet mixing angle, $\theta$; and doublet-singlet quartic coupling, $a_2$. Our key findings include:
    \begin{itemize}
    \item As illustrated in Figs.~\ref{fig:BM3_st_vByT} through \ref{fig:M600_vary_sintheta} and \ref{fig:sintheta-pheno},  for the heavy $M_2$ regime there exists a minimum value of $|\sin\theta|$ for which a first order EWPT occurs. For smaller values of $|\sin\theta|$ the transition is a smooth crossover. Importantly, this situation contrasts from the conclusion that one would infer from perturbative computations, which allow for a first order EWPT for arbitrarily small $|\sin\theta|$. The phenomenological implications are significant, particularly for precision studies of Higgs boson properties: future Higgs factories will access a greater portion of the first order EWPT-viable parameter space than previously inferred from one-loop perturbative results.

   \item Also in the heavy $M_2$ regime, two-loop perturbative computations agree remarkably well with the nonperturbative results for sufficiently strong transitions, ($\sqrt{2 \Delta \phi^\dag\phi}/T \gtrsim 1$). In many points the two-loop results are even good within $5\%$, not including perturbative uncertainty due to residual higher-order effects that we have not estimated.

    This behavior supports the intuitive picture that genuine nonperturbative effects related to the gauge sector become less important once the EWPT is strongly first-order. Perhaps surprisingly, perturbative convergence appears to be excellent despite our benchmark points involving a rather large coupling: $a_2 \sim 2.5 - 4.0$.
    
    \item
    Results from one-loop perturbative computations compare rather poorly with the nonperturbative results, and should, therefore, be considered a qualitative estimator of phase transition thermodynamics at best.
    Furthermore, use of perturbation theory alone in general never allows one to identify parameter space regions associated with crossover transitions, and even two-loop perturbative computations fail to yield the correct qualitative behavior for weakly first order transitions in the vicinity of the first order/crossover boundary.

    Given that the phase transition phenomenology in many BSM models is only understood at one-loop level, we believe our lattice results provide strong motivation for (a) extending these phenomenological analyses to at least the two-loop level and (b) comparing with lattice results for weak first order transitions.

    \item We also briefly discussed phase transitions with a light BSM scalar, in parameter space that carries phenomenological importance for precision studies of Higgs boson decays. We verified with simulations that a strongly first-order EWPT can occur in this parameter space. For this regime, the first order EWPT can occur at 
    considerably smaller portal couplings than for the heavy $M_2$ regime.

    Nevertheless, the transition can become crossover for sufficiently small $a_2$. We, thus, performed a perturbative two-loop scan of this region to map out strong transitions in the $(M_2, a_2)$ plane, extending the earlier one-loop work of ref.~\cite{Kozaczuk:2019pet}. As in the latter study, we find that the criteria that a transition occurs is decisive in setting a lower bound on $a_2$ and, thus, the exotic Higgs decay rate.
  \end{itemize}

    Simulation results for figures \ref{fig:BM3_st_vByT} through \ref{fig:small_mass_a2} is available on Zenodo \cite{NiemiZenodo:2024}.
    The simulation code is available at \url{https://github.com/niemilau/su2higgs}.

\vspace{0.1in}
	
	\begin{acknowledgments}
		We thank Oliver Gould, Kari Rummukainen, Tuomas V. I. Tenkanen and Yanda Wu for helpful discussions. LN was supported by Academy of Finland grants 320123, 345070 and 354572 and National Natural Science Foundation of China grant 11975150. MJRM and GX were supported under National Natural Science Foundation of China grants 11975150 and 12375094.
        We acknowledge CSC – IT Center for Science, Finland, for computational resources.
	\end{acknowledgments}

	\onecolumngrid
	\newpage
	\appendix
	\allowdisplaybreaks
	\begin{center}{\large \textbf{Appendix}} \end{center}

	\section{Lattice-continuum relations}
	\label{sec:lattice-counterterms}
	
	This Appendix collects the expressions needed to convert between 3D continuum and lattice actions, Eqs.~(\ref{eq:3D-EFT}) and (\ref{eq:lattice-act}). The 4D $\rightarrow$ 3D matching relations give the continuum parameters in \MSbar scheme, and in 3D the only parameters requiring renormalization are $\bar{b}_1, \bar{m}_\phi^2, \bar{m}_S^2$. Furthermore, they are divergent only up to two-loop order. We therefore write the bare lattice parameters as $m_{\phi,L}^2 = \bar{m}_\phi^2(\bar{\mu}) + \delta m_{\phi,L}^2$ etc, and fix the counterterm by computing a physical quantity in both schemes (here $\bar{\mu}$ is the \MSbar scale). The result is exact at two-loop order, up to corrections that vanish as $a\rightarrow 0$.
	
	Following Ref.~\cite{Laine:1995np}, the physical quantity of our choice is the value of the effective potential $\Veff$ in its minimum. The difference 
	\begin{align}
	\Delta \Veff = \Veff^\text{lattice} - \Veff^\text{\MSbar} 
	\end{align}
	is insensitive to IR physics and can thus be calculated safely in perturbation theory, as the two schemes can differ only in the UV region.
	If the fields are shifted as 
	\begin{align}
	\phi \rightarrow \phi + \varphi = \phi + \frac{1}{\sqrt{2}} \begin{pmatrix} 0 \\ v \end{pmatrix}, \quad\quad S \rightarrow S + s,
	\end{align}
	the required counterterms can be extracted directly from coefficients of $s, s^2$ and $v^2$ in $\Delta \Veff$. We also compute a vacuum counterterm $\delta V$ so that the mass-dependent part of the vacuum energy agrees in both schemes. $\delta V$ relates quadratic scalar condensates on lattice and in \MSbar through 
	\begin{align}
	\label{eq:condensate-renormalization}
	\langle\he\phi\phi \rangle_\text{\MSbar} = \langle\he\phi\phi \rangle_L + \frac{\partial (\delta V)}{\partial \bar{m}_\phi^2}, \quad\quad \langle S^2 \rangle_\text{\MSbar} = \langle S^2 \rangle_L + 2\frac{\partial (\delta V)}{\partial \bar{m}_S^2}
	\end{align}
	where the lattice condensates are computed in the theory without $\delta V$ (i.e. Eq.~\ref{eq:lattice-act}). The linear singlet condensate $\langle S \rangle$ does not require renormalization.
	
	The calculation of $\Delta \Veff$ proceeds largely as outlined in Refs.~\cite{Laine:1995np, Laine:1997dy}, from where one can find all required loop integrals and lattice vertices associated with the $\SU2$ + Higgs sector (in Feynman-'t Hooft gauge with $\xi = 1$). Extension to the $\SU2 + \grU1$ case is straightforward, and the singlet field brings no qualitatively new features. We note that in the two-loop part, we may neglect quadratic mixing among the different fields and work with undiagonalized propagators, somewhat simplifying the calculation. This is possible because two-point vertices that mix propagators are associated with reduced UV sensitivity and contribute only $\mathcal{O}(a)$ corrections to the $\Veff$ difference. Note also that expanding the $\grU1$ lattice action produces a four-photon vertex whose contribution to $\Delta \Veff$ is $-\bar{g}'^4 r^2 \Sigma v^2 T^2 / (192\pi)$, where $\Sigma$ is given in Eq.~(\ref{eq:lattice-constants}), plus an irrelevant vacuum divergence.

	We omit further details of the calculation as it is a straightforward application of the methods described in Refs.~\cite{Laine:1995np, Laine:1997dy}. The resulting expressions for lattice parameters read 
	\begin{align}
	\label{eq:lattice-ct-start}
	m_{\phi,L}^2 &= \bar{m}_\phi^2(\bar{\mu}) - \frac{\Sigma T}{8\pi a}\left(3 \bar{g}^2 + \bar{g}'^2 + 12 \bar{\lambda} + \bar{a}_2  \right) \nn 
	&\; + \frac{T^2}{(4\pi)^2} \Big[ \Big( -\frac{51}{16} \bar{g}^4 + \frac{5}{16} \bar{g}'^4 + \frac98 \bar{g}^2 \bar{g}'^2 - 3 (3\bar{g}^2 + \bar{g}'^2) \bar{\lambda} + 12 \bar{\lambda}^2 + \frac12 \bar{a}_2^2 \Big)\Big( \ln\frac{6}{a \bar{\mu}} + \zeta \Big)  \nonumber \\
	& + 3 \bar{\lambda} \left( 3\bar{g}^2 + \bar{g}'^2 \right) \left(\delta - \frac14 \Sigma^2 \right) + \bar{g}^4 \Big( -\frac{15}{16} - \frac{45}{64}\Sigma^2 - \frac{\pi}{4}\Sigma + \frac{33}{8}\delta + \frac{9}{2}\rho - 3\kappa_1 + \frac{3}{2}\kappa_4 \Big) \nn 
	& \; + \bar{g}'^4 \Big( \frac{1}{16} - \frac{1}{64}\Sigma^2 - \frac{\pi r^2}{6}\Sigma + \frac{1}{8}\delta + \frac12 \rho \Big) + \bar{g}^2 \bar{g}'^2 \Big( \frac38 - \frac{3}{32} \Sigma^2 + \frac34 \delta \Big)
	\Big] \\
	%
	%
	\nn
	m^2_{S,L} &= \bar{m}_S^2(\bar{\mu}) - \frac{\Sigma T}{4\pi a} (2 \bar{a}_2 + 3 \bar{b}_4) \nn 
	& \; + \frac{T^2}{(4\pi)^2} \Big[ \Big( 2 \bar{a}_2^2 + 6 \bar{b}_4^2 - \bar{a}_2 \left( 3\bar{g}^2 + \bar{g}'^2\right) \Big)\Big( \ln\frac{6}{a \bar{\mu}} + \zeta \Big) + \bar{a}_2 \left(3\bar{g}^2 + \bar{g}'^2 \right) \left(\delta - \frac{1}{4}\Sigma^2\right) \Big] \\
	%
	%
	\nn
	\label{eq:lattice-ct-end}
	b_{1,L} &= \bar{b}_1(\bar{\mu}) - \frac{\Sigma T}{4\pi a}\left(\bar{a}_1 + \bar{b}_3 \right) \nn 
	& \; + \frac{T^2}{(4\pi)^2} \Big[ \Big( \bar{a}_1 \bar{a}_2 + 2 \bar{b}_3 \bar{b}_4 - \frac{\bar{a}_1}{2} \left( 3\bar{g}^2 + \bar{g}'^2 \right) \Big) \Big( \ln\frac{6}{a \bar{\mu}} + \zeta \Big) + \frac{\bar{a}_1}{2} \left( 3\bar{g}^2 + \bar{g}'^2 \right) \left( \delta - \frac{1}{4}\Sigma^2 \right) \Big].
	\end{align}
	The vacuum counterterm is 
	\begin{align}
	\delta V &= -\frac{\Sigma T}{4\pi a} \left( \frac12 \bar{m}_S^2 + 2 \bar{m}_\phi^2 \right) + \frac{T^2}{(4\pi)^2} \left( 3\bar{g}^2 + \bar{g}'^2 \right) \bar{m}_\phi^2 \Big[ - \Big( \ln\frac{6}{a \bar{\mu}} + \zeta \Big) + \left(\delta - \frac14 \Sigma^2 \right) \Big] \nn 
	& \; + (\text{independent of masses and } \bar{b}_1).
	\end{align}
	We stress once more that because of super-renormalizability, these relations are exact in the $a\rightarrow 0$ limit. As a cross-check, we may decouple the singlet to find agreement with the $\SU2 \times \grU1$ + Higgs counterterms given in \cite{Laine:1997dy} (in their notation $\gamma = r^{-1}$). The pure singlet limit agrees with \cite{Gould:2021dzl}.

	Various constants appearing in the counterterms above are \cite{Laine:1997dy}
	\begin{align}
	\label{eq:lattice-constants}
	\Sigma &= 3.175911535625, \quad\quad \delta = 1.942130(1), \quad\quad \rho = -0.313964(1) \nn
	& \zeta = 0.08849(1), \quad\quad \kappa_1 = 0.958382(1), \quad\quad \kappa_4 = 1.204295(1).
	\end{align}
	These quantities originate from numerical evaluation of loop integrals on the lattice. The parentheses represent uncertainty in the last digit.

	\section{Reweighting}
	\label{sec:reweight}
	
	Suppose we run a simulation at fixed temperature $T$ and obtain the probability distribution (normalized histogram) of an observable $A$, $p_T(A)$. Its functional form is 
	\begin{align}
	p_T(A') \propto \int [d\varphi] \; \delta(A(\varphi) - A') e^{-S(T)},
	\end{align}
	where $\varphi$ collectively denotes all fields in the action and $S(T)$ is the 3D lattice action from Eq.~(\ref{eq:lattice-act}) evaluated at temperature $T$.
	The corresponding distribution at any other temperature $T'$ can be written as
	\begin{align}
	p_{T'}(A') \propto \int [d\varphi] \; \delta ( A(\varphi) - A') e^{-S(T')} = \int [d\varphi] \; \delta (A(\varphi) - A') W_{T,T'} e^{-S(T)},
	\end{align}
	where 
	\begin{align}
	\label{eq:rw-factor}
	W_{T,T'} = e^{-\left( S(T') - S(T) \right)}
	\end{align}
	is the reweighting factor. The histogram at temperature $T'$ can therefore be obtained using data from a simulation performed at a different temperature $T$ by weighting the measurements of $A$ by the factor $W_{T,T'}$. This is the basic idea of histogram reweighting \cite{Ferrenberg:1988yz}. We demonstrate below how the temperature dependence of the action can be expressed in terms of easily measurable quantities, providing a convenient way of computing the reweighting factor.

	To make use of reweighting, we must know $W_{T,T'}$ for each field configuration from which measurements are taken. The temperature dependence of the lattice action is complicated because the 4D $\rightarrow$ 3D mapping makes all 3D parameters $T$ dependent. Their temperature dependence can nevertheless be found by following the steps described in sections~\ref{sec:EFT} and \ref{sec:lattice-theory}. The factor $W_{T,T'}$ can then be constructed from volume averages of local operators appearing in the action. In practice it is useful to scale out the lattice spacing $a$ and the temperature by defining
	\begin{gather}
	\hat{\phi} = \sqrt{aT^{-1}}\phi, \quad \hat{S} = \sqrt{aT^{-1}}S, \nn
	\hat{m}_{\phi,L}^2 = a^2 m_{\phi,L}^2, \quad \hat{m}_{S,L}^2 = a^2 m_{S,L}^2, \quad \hat{\lambda} = aT \bar{\lambda}, \quad \hat{b}_4 = aT \bar{b}_4, \quad \hat{a}_2 = aT \bar{a}_2, \nn
	\hat{b}_{1,L} = T^{-1/2} a^{5/2} b_{1,L}, \quad \hat{b}_3 = \sqrt{T} a^{3/2} \bar{b}_3, \quad \hat{a}_1 = \sqrt{T} a^{3/2} \bar{a}_1
	\end{gather}
	and perform a change of field variables to $\hat\phi, \hat{S}$ in the functional integral.

	\begin{figure*}[!htb]
		\subfloat{\includegraphics[width=0.5\textwidth]{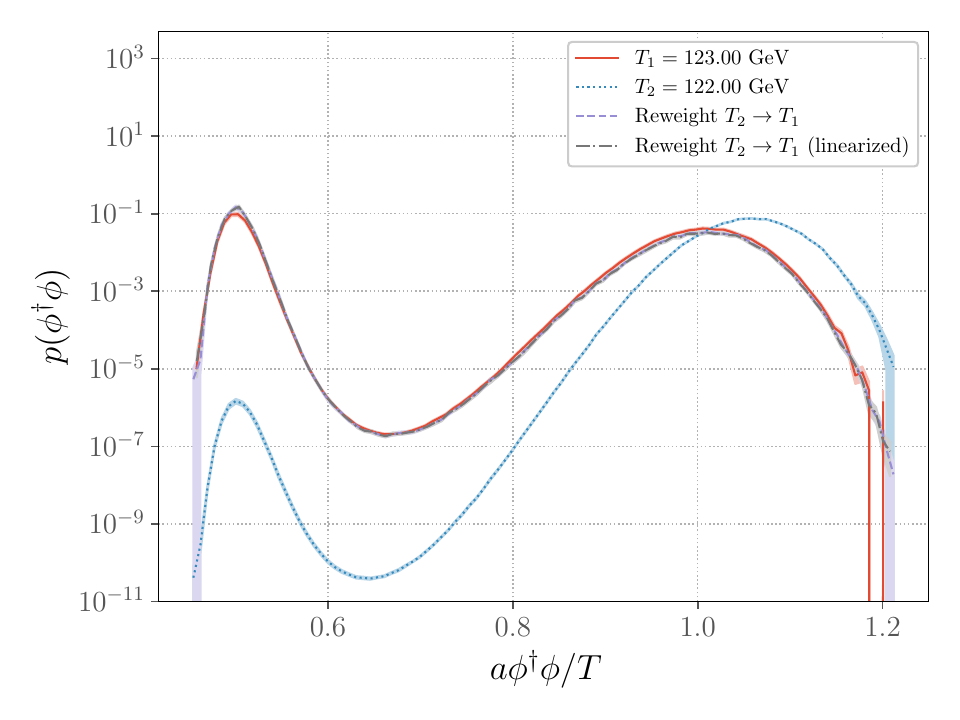}}
		\subfloat{\includegraphics[width=0.5\textwidth]{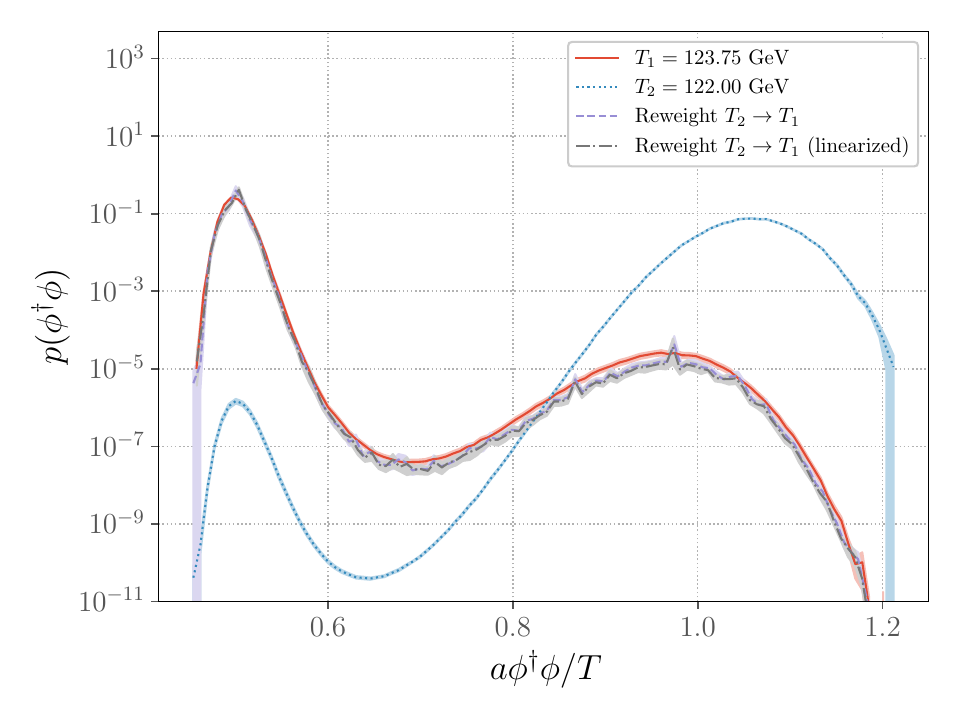}}
		\caption{
			Testing our reweighting routine in the benchmark point (\ref{eq:BM3}) with $\beta = 12, N_s = 18$. We have performed separate simulations at temperatures, $T_1$ being close to the critical temperature, then reweighted the histogram of $\he\phi\phi$ from $T_2$ to $T_1$ (at $T_2$, the input $\beta$ is adjusted so that the spacing $a$ has same physical value as at $T_1$). Left-hand plot has $|T_2 - T_1| = 1$ GeV while the plot on the right uses a more aggressive reweighting range of $|T_2 - T_1| = 1.75$ GeV. Statistical errors are shown by shaded bands. As expected, the reweighted histogram(s) agree well with the original histogram at $T_1$, although the right-hand plot shows some loss in accuracy due to the larger temperature difference. Curves labeled ``linearized'' are reweighted using the approximate formula Eq.~(\ref{eq:rw-factor-linearized}) instead of Eq.~(\ref{eq:rw-factor}); difference between the two methods is negligible.
		}
		\label{fig:reweight-test}
	\end{figure*}
	
	Denoting volume-averaged operators as $(O)_V \equiv a^3 \sum_x O(x) / V$, the lattice action is 
	\begin{align}
	\label{eq:action-volume-avg}
	a^3 S_L(T) / V  = \; & \beta(T) \left( \sum_{i<j} \left[ 1 - \frac{1}{2} \RE \Tr P_{ij}(x) \right] \right)_V + \beta'(T) \left( \sum_{i<j} \Big[ 1 - \RE p_{ij}^{r}(x) \Big] \right)_V \nn  
	& + \hat{m}_{\phi,L}^2(T) (\he{\hat\phi}\hat\phi)_V + \hat{\lambda}(T) (\he{\hat\phi}\hat\phi)^2_V + \hat{b}_{1,L}(T) (\hat{S})_V + \frac12 \hat{m}_{S,L}^2(T) (\hat{S}^2)_V + \frac13 \hat{b}_3(T) (\hat{S}^3)_V \nn 
	&  + \frac14 \hat{b}_4(T) (\hat{S}^4)_V  + \frac12 \hat{a}_1(T) (\hat{S} \he{\hat\phi}\hat\phi)_V + \frac12 \hat{a}_2(T) (\hat{S}^2 \he{\hat\phi}\hat\phi)_V + \text{(scalar kinetic terms)}.
	\end{align}
	Our simulation program separately measures and stores the volume averages appearing in this expression. It is then straightforward to calculate the reweighting factors $W_{T,T'}$ in post process stage without having to store the field configurations.
	In the parametrization used here, the lattice scalar kinetic terms have no explicit $T$-dependence and do not contribute to $W_{T,T'}$.
	
	For numerical analysis we find it useful to expand the difference $S(T') - S(T)$ in small  $(T' - T) / T$ and approximate 
	\begin{align}
	\label{eq:rw-factor-linearized}
	W_{T,T'} \approx \exp[ -S'(T) (T' - T) ].
	\end{align}
	We utilize this ``linearized'' form over the full expression (\ref{eq:rw-factor}) throughout the paper as it was simpler to implement in our existing tool set. This is an excellent approximation: In Fig.~\ref{fig:reweight-test}, we show reweighting of example order parameter histograms from $T_2$ to $T_1$ using either (\ref{eq:rw-factor}) or (\ref{eq:rw-factor-linearized}). The difference between the two methods is almost indistinguishable.
	
	\bibliographystyle{apsrev4-1}
	\bibliography{singletrefs}
	
\end{document}